\documentclass[fleqn,usenatbib,useAMS]{mnras}

\usepackage{graphicx}	
\usepackage{amsmath}	
\usepackage{amssymb}	
\usepackage{multicol}        
\usepackage{bm}		
\usepackage{pdflscape}	
\usepackage{longtable}

\usepackage{hyperref}

\usepackage{natbib}
\setcitestyle{authoryear,open={(},close={)}}

\newcommand{\Rha}{$R\mathrm{_{h}}=0.6 \,\mathrm{pc}\,$}
\newcommand{\Rhb}{$R\mathrm{_{h}}=1.0 \,\mathrm{pc}\,$}

\newcommand{\W}{$W\mathrm{_{0}}\,$}

\newcommand{\f}{$f\mathrm{_c}\,$}
\newcommand{\fa}{$f\mathrm{_c}=0.1\,$}
\newcommand{\fb}{$f\mathrm{_c}=0.5\,$}
\newcommand{\fc}{$f\mathrm{_c}=1.0\,$}

\newcommand{\msun}{$\mathrm{M_{\odot}}\,$} 

\usepackage[T1]{fontenc}
\usepackage{ae,aecompl}

\usepackage{newtxtext,newtxmath}

\title[MNRAS \LaTeX\ guide for authors]{Intermediate Mass Black Hole Formation in compact Young Massive Star Clusters}
\author[F. P. Rizzuto]{Francesco Paolo Rizzuto$^{1}$\thanks{Contact e-mail: \href{mailto:}{rizzuto@MPA-Garching.MPG.DE} },
Thorsten Naab$^{1}$, Rainer Spurzem$^{2,3,4}$\thanks{Research Fellow at Frankfurt Institute for Advanced Studies
(FIAS)}, Mirek Giersz$^{5}$, \newauthor J. P. Ostriker$^{6,7}$, N. C. Stone$^{6, 8, 9}$, Long Wang$^{11,12}$, 
Peter Berczik$^{2,3,10}$\thanks{ORCID: \href{mailto:}{https://orcid.org/0000-0003-4176-152X} },
M. Rampp$^{13}$,  
\\ %
$^{1}$Max-Planck Institute for Astrophysics, Karl-Schwarzschild-Str. 1,D-85741 , Garching, Germany
\\
$^{2}$National Astronomical Observatories and Key Laboratory of Computational Astrophysics, Chinese Academy of Sciences,20A Datun Rd., 
\\
Chaoyang District, 100101, Beijing, China
\\
$^{3}$Astronomisches Rechen-Institut, Zentrum für Astronomie, University of Heidelberg, Mönchhofstrasse 12-14,69120, Heidelberg, Germany
\\
$^{4}$Kavli Institute for Astronomy and Astrophysics, Peking University, Yiheyuan Lu 5, Haidian Qu, 100871, Beijing, China
\\
$^{5}$Nicolaus Copernicus Astronomical Centre, Polish Academy of Sciences, ul. Bartycka 18, 00-716 Warsaw, Poland
\\
$^{6}$Department of Astronomy, Columbia University, New York, NY 10027, USA
\\
$^{7}$Department of Astrophysical Sciences, Princeton University, Princeton, NJ 08544, USA
\\
$^{8}$Racah Institute of Physics, The Hebrew University, Jerusalem, 91904, Israel
\\
$^{9}$Department of Astronomy, University of Maryland, College Park, MD, 20742, USA
\\
$^{10}$Main Astronomical Observatory of Ukrainian National Academy of Sciences, Kiev, Ukraine 
\\
$^{11}$Argelander Institut für Astronomie, Auf dem Hügel 71, Bonn, Germany
\\
$^{12}$RIKEN Center for Computational Science 7-1-26 Minatojima-minami-machi, Chuo-ku, Kobe, Japan
\\
$^{13}$Max Planck Computing and Data Facility, Gießenbachstr. 2, Garching, Germany
}

\date{\today}

\pubyear{2020}

\begin{document}
\label{firstpage}
\pagerange{\pageref{firstpage}--\pageref{lastpage}}
\maketitle

\begin{abstract}
Young dense massive star clusters are a promising environment for the formation of intermediate mass black holes (IMBHs) through collisions. We present a set of 80 simulations carried out with Nbody6++GPU of 10 initial conditions for compact $\sim 7 \times 10^4 M_{\odot}$ star clusters with half-mass radii $R_\mathrm{h} \lesssim 1 pc$, central densities $\rho_\mathrm{core} \gtrsim 10^5 M_\odot pc^{-3}$, and resolved stellar populations with 10\% primordial binaries. Very massive stars (VMSs) with masses up to $\sim 400 M_\odot$ grow rapidly by binary exchange and three-body scattering events with main sequences stars in hard binaries. Assuming that in VMS - stellar BH collisions all stellar material is accreted onto the BH, IMBHs with masses up to $M_\mathrm{BH} \sim 350 M_\odot$ can form on timescales of $\lesssim 15$ Myr. This process was qualitatively predicted from Monte Carlo MOCCA simulations. Despite the stochastic nature of the process - typically not more than 3/8 cluster realisations show IMBH formation - we find indications for higher formation efficiencies in more compact clusters. Assuming a lower accretion fraction of 0.5 for VMS - BH collisions, IMBHs can also form. The process might not work for accretion fractions as low as 0.1. After formation, the IMBHs can experience occasional mergers with stellar mass BHs in intermediate mass-ratio inspiral events (IMRIs) on a 100 Myr timescale. Realised with more than $10^5$ stars, 10 \% binaries, the assumed stellar evolution model with all relevant evolution processes included and 300 Myr simulation time, our large suite of simulations indicates that IMBHs of several hundred solar masses might form rapidly in massive star clusters right after their birth while they are still compact.
\end{abstract}
\begin{keywords}

gravitational waves – methods: numerical – stars: black holes – stars: dynamics – stars: mass-loss – galaxies: star clusters: general.

\end{keywords}



\begingroup
\let\clearpage\relax
\endgroup
\newpage

\section{Introduction}



At the end of the 1930s \cite{Oppenheimer1939} suggested that after the exhaustion of all thermonuclear energy sources,  heavy stars might collapse. The objects created in these events are stellar black holes (BH) with masses from $5$ \msun up to about $60$ \msun \footnote{The lower and upper mass boundaries depend on the stellar evolution model, i.e. details of stellar wind mass loss and supernova explosions}. 
A large number of  X-ray and optical observations provide solid evidence for the existence of stellar BHs \citep{Webster1972,Remillard2006,Casares2014}. Their presence is further confirmed by the recent discovery of gravitational waves generated by BH mergers 
\citep{Abbott2016,Abbott2017, Lange2018}. It is also well established that massive galaxies in the local Universe host supermassive black holes (SMBHs) with masses above $10^6$ \citep[see][for a general review]{Kormendy2013}. However, there is very little observational evidence for the existence of BHs bridging the mass range between stellar and supermassive black holes. These intermediate-mass black holes (IMBHs) could originate from stellar BHs and might be the seeds for SMBHs. Finding them and understanding their formation mechanism is crucial for a full understanding of the BH population in the Universe.  
There are three theoretical paths for IMBH formation leading to SMBHs discussed in the literature \citep[see reviews][and citations there]{Volonteri2010, Koliopanos2017}.

In the first scenario IMBHs form through direct collapse of dense gas at high redshifts. 
This scenario predicts the formation of IMBHs of $10^4$ to $10^6$ \msun \citep{Begelman2006, Agarwal2012, Luo2020}. A second possibility is that IMBHs are the remnants of first generation (PopIII) stars. These stars formed from zero metallicity gas and are expected to collapse into  IMBHs more massive than $100$ \msun \citep{Madau2001, Ryu2016}. A third family of models assumes that IMBHs are generated in dense stellar environments through runaway collisions. Several studies have demonstrated that IMBHs can form through dynamical interactions in dense stellar systems. Those studies include analytical approaches \citep{Begelman1978, Stone2017} as well as N-body simulations \citep{Zwart2002, Zwart2004, Mapelli2016, DiCarlo2020}, and Monte Carlo simulations \citep{Freitag2006,  Gurkan2006, Giersz2015}. In particular, the effect of tidal capture of stars by BHs has been discussed in \citet{Patruno2006} for massive star clusters and \citet{Stone2017} for nuclear star clusters.

Several IMBHs candidates have been discovered in our galaxy and others nearby. For example, IMBHs have been proposed to explain the nature of ultra-luminous X-ray emitters (ULXs). ULXs are extra-galactic and off-center X-ray sources that could be generated by BHs of intermediate mass which accrete gas isotropically below the Eddington rate \citep{Colbert1999}. Most ULXs observed are, however, more likely generated by smaller objects such as magnetized neutron stars and stellar BHs with super-Eddington accretion \citep{Feng2011, Gladstone2013, Roberts2016, Kaaret2017, King2020}. This is also confirmed by dynamical evidence \citep{Liu2013} as well as X-ray pulsations, which indicate the presence of neutron stars \citep{Walton2016}. Nevertheless, there exists a group of hyper-luminous X-ray sources (HLXs)  that might be laborious to explain by super-Eddington accretion since their luminosity exceeds $ \sim10^{41} $ erg s$^{-1}$. Probably the best IMBH candidate known so far is HLX-1. This HLX is believed to host an IMBH because  it has an X-ray luminosity of $1.1 \times 10^{42}$ erg s$^{-1}$ \citep{Farrell2009}. 
This luminosity would imply a mass of $500$ \msun  even assuming an accretion rate ten times larger than the Eddington limit  \citep{Farrell2009}. 
The mass estimates of the BH associated with HLX-1 is estimated between $3.0 \times 10^3$ and $3.0 \times 10^5$ \msun \citep[see review][and references therein]{Mezcua2017}.  Other notable  sources in the same category are NGC 5252 and NGC 2276-3C, which have been estimated to host IMBHs of $\sim 10^5 M_{\odot}$  \cite{Kim2020} and $\sim 5 \times 10^4 M_{\odot}$ \citep{Mezcua2013, Mezcua2015}, respectively. 
Another HLX is M82 X-1.  This X-ray source is associated with a young massive star cluster (MGG-11) in the starburst galaxy M82 \citep{Matsumoto1999, Kaaret2001}. First observations have suggested that it is generated by a BH with a mass of $200-5000$ \msun \citep{Kaaret2001,Matsumoto2001,Strohmayer2003,Patruno2006}. A more recent analysis indicates the presence of an IMBH, estimating its mass to $\sim 400$ \msun \citep{Pasham2014}. However, M82 X-1  could still be a stellar mass BH with super-Eddington accretion \citep{Brightman2016}.   

Globular clusters have been popular targets for the search of IMBHs. Due to their high central density, and possibly even higher density at formation \citep{2019ApJ...879L..18L}, they provide a promising environment for the formation of IMBHs through runaway core-collapse and collision. Many studies have attempted to detect IMBHs in globular clusters through their accretion signatures. However, so far G1 in M31 is the only globular cluster detected in X-rays \citep{Pooley2006, Kong2007}. This signal might be generated by an IMBH with a mass of about $2 \times 10^4$ \msun \citep{Gebhardt2002, Gebhardt2005}. 
A recent  observational study reports the lack of  IMBHs  accretion signatures in 19 globular clusters located in the  early-type galaxy NGC 3115 \citep{Wrobel2020}.
Here it is important to note that globular clusters contain little gas, which might explain the absence of X-ray and radio signals. Observations based on kinematic measurements might suggest the presence of IMBHs in globular clusters such as  M15 \citep{Bahcall1976, Peterson1989}, $\omega$ Centauri \citep{Noyola2008, Noyola2010}, NGC 1904 and  NGC 6266  \citep{Luetzgendorf2013}. However, these observations could also be explained by a central concentration of compact objects \citep{Baumgardt2003, vandenBosch2006, Baumgardt2020}. Measurements of pulsar accelerations indicate that the globular cluster Tucanae 47 might host an IMBH of $2300^{+1500}_{-850}$ \msun \citep{kizi2017}. Another study suggests, that current observations of pulsar accelerations are insufficient to confirm the presence of an IMBH in Tucanae 47, they can only be used to estimate an upper limits on its mass \cite{Abbate2019}.

Dwarf galaxies are very promising systems for IMBH searches. 
NGC 4395 seems to be one of the most plausible candidates for an active and central IMBH. 
Observations of the central stellar velocity dispersion reveal a value of $30$ km/s \citep{Filippenko2003}, 
suggesting an IMBH mass of $\sim 10^5$ \msun. Further measurements based on the broad profile of the H$\beta$ line, from X-ray variability \citep{Filippenko2003}, reverberation mapping \citep{Peterson2006,Haim2012}, and integral field kinematics \citep{denBrok2015} indicate a mass in the range between $10^4$ to $10^5$ \msun.
Also our Galaxy might host an IMBH in the vicinity of  the galactic center as suggested by recent high-resolution molecular line observations that indicates the presence of a $\sim 10^4$ \msun candidate BH in the central region of the  Milky Way \citep{Takekawa2020}.

In this paper, we study IMBH formation paths in young concentrated star clusters with numerical simulations. The simulations incorporate detailed models for single and binary stellar evolution as well as mass loss due to stellar winds. In Section 2, we describe the N-body code used to simulate star cluster evolution and focus on the description of the adopted stellar evolution and collision models. In Section 3, we describe the initial conditions for our models. The results of our simulations are discussed in Section 4. In Section 5 we  compare these results with previous studies. In the final section we summarise the main points of this paper.




\section{The Method}
 \begin{table*}
  \begin{tabular}{lccccccccclll}
    \hline
  Model Name       & $r\mathrm{_c}\,$      & $\rho_c$   & \W & $R\mathrm{_h}\,$ & $\sigma$ & $t_{\mathrm{rh}}$ & $t_{\mathrm{s}}$  & \f   & \# IMBH   &  $M_{\mathrm{IMBH}}$ & $t_{\mathrm{form}}$  & $M_*/M_{\mathrm{IMBH}}$     \\
        & pc    & \msun / pc $^3$   &       & pc & km/s   &    Myr & Myr  &                    &            & \msun & Myr &  \\
  \hline
  R06W9F01  & $0.04$ & $3.0\times10^7$ & $9$   & $0.6$ & 15 & 56   &1.4& $0.1$ & $0/8$  & /           & / & /   \\
  R06W9F05  & $0.04$ & $3.0\times10^7$ & $9$   & $0.6$ & 15 & 56   &1.4& $0.5$ & $2/8$  & 138, 110                  &5.5,   6.3                        & 79\%,  81\%,  \\ 
  R06W6        & $0.19$ & $1.1\times10^5$ & $6$   & $0.6$ & 15 & 56   &1.4& $1.0$ & $4/8$  & 307, 151, 138, 122   &8.6,  83.9,  6.4,  8.4      &  86\%,  26\%,  72\%,  80\%\\ 
  R06W7        & $0.13$ & $5.0\times10^5$ & $7$   & $0.6$ & 15 & 56   & 1.4&$1.0$ & $2/8$  & 148, 147                   &22.9 , 8.2             		  & 76\%, 87\% \\            
  R06W8        & $0.06$ & $3.0\times10^6$ & $8$   & $0.6$ & 15 & 56   & 1.4&$1.0$ & $3/8$  & 336, 171,110            &6.6, 12.3, 8.1        		  & 98\%, 87\%, 77\% \\  
  R06W9        & $0.04$ & $3.0\times10^7$ & $9$   & $0.6$ & 15 & 56   & 1.4&$1.0$ & $3/8$  & 355, 349, 120           &8.57, 8.19, 16.3     	    & 81\%,  89 \%,  73 \%\\   
  R1W7          & $0.2$ & $1.0\times10^5$ & $7$   & $1.0$ &   12 &120   & 3.0&$1.0$ & $0/8$  & /                               & /                                  &   /                             \\
  R1W8          & $0.11$ & $4.0\times10^5$ & $8$   & $1.0$ & 12 &120   & 3.0&$1.0$ & $0/8$  & /                               & /                                  &   /                 \\
  R1W9          & $0.05$ & $3.0\times10^6$ & $9$   & $1.0$ & 12 & 120   & 3.0&$1.0$ & $1/8$  & 239                           &16.4                               & 92\%                 \\   
  R1W10        & $0.03$ & $1.8\times10^7$ & $10$  & $1.0$ & 12 & 120  & 3.0&$1.0$ & $2/8$  & 133, 110                  & 13.2, 7.8                       & 83\%,  83 \%             \\ 
  \end{tabular}
  \caption{Model parameters of the cluster simulations: $r\mathrm{_c}\,$: initial core radius; $\rho\mathrm{_c}\,$: initial central density; \W: central potential parameter for the King density profile \citep{King1966}; $R\mathrm{_h}\,$: half mass radius; $\sigma$ : dispersion velocity;  $t_{\mathrm{rh}}$: half mass relaxation time computed using eq. \ref{eq:trh} with $\gamma=0.02$; $t_{\mathrm{s}}$: segregation time scale for $100$ \msun objects computed using eq. \ref{eq:ts}; \f: fraction of mass absorbed by a compact object during a direct collision with a star; \# IMBH : Number of  BHs with masses $\geq 100$ \msun formed out of 8 realisations; $M_{\mathrm{IMBH}}$: IMBH masses; $t_{\mathrm{form}}$: IMBHs formation times; $M_*/M_{\mathrm{IMBH}}$: the total stellar mass accreted by the IMBH divided by its final mass.}
   \label{table:initial}
\end{table*}

To investigate the possible formation of IMBHs in massive star clusters we generated initial conditions for $80$ isolated systems\footnote{We study clusters in isolation to investigate internal dynamical effects without possible external influences.} using  MCLUSTER \citep{mcluster}. The systems were set up with two different half-mass radii and various central concentrations. Each initial condition is evolved  for a few hundred million years employing NBODY6++GPU \citep{Wang2015,Wang2016}, a direct N-body simulation code designed to follow the dynamical and stellar evolution of individual stars and binaries.

\subsection{NBODY6++GPU}

NBODY6++GPU\footnote{Link to repository: http://silkroad.bao.ac.cn/repos/Nbody6++GPU-Aug2020/} is a high-precision direct N-body simulation code based on the earlier N-body codes NBODY1-6 \citep{Aarseth1999} and NBODY6++ \citep{Spurzem1999}. It uses for time integration Taylor series up to 4$^{\rm th}$ order; due to the Hermite scheme it can be based on two time points only. This together with the hierarchically blocked variable time step scheme allows an efficient parallelization of the code for massively parallel supercomputers (since NBODY6++); gravitational forces between particles are offloaded to graphics processing units (GPUs), used for high-performance general purpose computing (NBODY6++GPU, \cite{Wang2015}). The parallelisation is achieved via MPI and OpenMP on the top level, distributing work within a group of particles due for time integration, and efficient parallel use of GPU cores at the base level (every MPI process using a GPU), for computing the gravitational forces between particles. 
The GPU implementation in NBODY6++GPU provides a significant performance improvement, especially for the long-range (regular) gravitational forces \citep[see][]{nitadori2012,Wang2015,Wang2016}.

The code accurately computes the evolution of binaries, multiples and close encounters between them and single stars and between multiple systems, using the Kustaanheimo-Stiefel  \citep[KS,][]{KS1965} regularisation with the classical chain algorithm by \citet{mikkola1998}. It also follows single and binary stellar evolution based on the SSE and BSE recipes by Hurley (see Sect. \ref{sec:stellarEvolution}), including also rapid tidal circularization for binaries with small pericenters and tidal captures according to the prescription given in \citet{Mardling2001}, which is based on the previous work of \citet{Press1977, LeeandOstriker1986, Mardling1996}.
The integrator fully resolves orbits and dynamical evolution of binaries, even during phases of mass loss or when one of the two stars undergoes a supernova explosion. The binary orbit is adjusted to the corresponding loss of mass, energy and angular momentum with appropriate time stepping; in case of a supernova explosion it is always ensuring that the remnant and its companion leave the explosion with the corrected orbital positions and velocities.

In the implementation used for this study we compute the gravitational wave energy loss of hard binaries according to the orbit averaged approximation of \citet{Peters1963}, using the average change of energy and angular momentum per orbit from their work. At each KS integration time-step, which is much smaller than the orbital time, we apply a corresponding fractional loss of energy and angular momentum. This allows for the proper representation of the evolution of gravitational radiation driven shrinking and circularization of the orbit, until the time scale of orbit shrinking becomes comparable to the orbital time. When this happens, the time to final coalescence is very short and we assume coalescence.

 The publicly available code NBODY6++GPU has been significantly upgraded in three respects:
\begin{enumerate}
\item For collisions between a compact remnant and a main sequence star or red giant a free parameter $f_c$ is introduced, which described the mass loss from the system in the process. The previous NBODY6 versions used only $f_c=1$, i.e. no mass loss in the process (see Eq. (1) and (2) above). Routine involved: {\tt coal.f}.
\item Simultaneous treatment of classical tidal interactions (Roche lobe overflow) and Post-Newtonian orbit-averaged orbit shrinking due to gravitational wave emission has been made possible. Both are treated technically in a similar way, and can now be switched on together. Routines and parameters involved: {\tt ksint.f, kstide.f, tides3.f, KZ(27)}.
\item Strongly bound binaries of two compact objects, which are subject to Post-Newtonian relativistic energy loss are prevented from unperturbed two-body integration. Routine involved: {\tt unpert.f}.
\end{enumerate}




\subsection{Stellar Evolution}

\label{sec:stellarEvolution}

The simulations in this paper were performed with the same stellar evolution models as the DRAGON simulations presented in \citet{Wang2016}. Stellar evolution is implemented using analytical fits to the models of \citep{Eggleton1989, Eggleton1990} developed by \citep{Hurley2000} and \citep{hurley2001} for single stars (SSE) and by  \citep{hurley2002} for binary stars (BSE). A few updates were included for strong kicks at neutron star birth \citep{hobbs2005} and for fallback and more massive black hole formation of massive stars at low metallicities \citep{Belczynski2002}.  

The code is able to follow the main properties of single stars (such as radius, mass, luminosity, and core mass) from the zero-age main-sequence to the remnant stage. This also includes mass loss due to stellar winds for a wide range of masses and metallicities.
As long as the orbit of a binary star is wide enough, the evolution of each star is assumed not to be affected by its companion and just the single star tracks are used. However, if one of the two stars is losing mass by a stellar wind, the companion has the chance to accrete material and deviate from its standard evolution. For close enough orbits either star might fill its Roche-lobe leading to mass transfer. For these cases, the code computes the accretion rate as a function of the masses, the radii, the stellar types, and the separation of the donor and the accretor, ensuring that it never exceeds 100 times the Eddington limit. If the matter ejected by a star is not entirely absorbed by its companion it might accumulate in a common envelope around the two stars. All the above effects: mass transfer, Roche phase, and common envelope evolution have significant consequences for the orbit and stellar properties of the binary and are included in the simulations based on the models of  \cite{Tout1997}.

In dense environments, where stellar collision rates are high, runaway collisions \citep{Lee1987, Quinlan1987, Quinlan1989, Quinlan1990},  can generate stars  above the maximum IMF mass of $100$ \msun \citep{Zwart2002, Zwart2004,Gurkan2004,Mapelli2016,DiCarlo2020,Wang2020} for which we use the term "very massive stars" (VMS). To track the evolution of VMSs, in the absence of observational constraints, we extrapolate our stellar evolution model to stars with arbitrary large masses. Therefore, VMSs are affected by strong stellar wind mass loss and lose a significant fraction of their mass during their lifetime.  In our stellar evolution framework, it is therefore impossible to form massive BHs from direct stellar collapse. In fact, with the adopted stellar evolution recipes, even an isolated star with a mass of $500$ \msun generates a BH of only about $30$ \msun. In the absence of well-established theories for VMSs evolution,  we therefore assume a conservative model \footnote{Our recipes do not include pair-instability supernovae and pulsation pair instability.}.
It is worth mentioning that theories of stellar evolution predict that a low metallicity star more massive than $260$ \msun collapses directly into a BH  without significant mass loss in supernova explosions \citep[see][and citations therein]{Woosley2015}. However, it is important to take into account the complex stellar structure acquired by  VMSs during merger events. A detailed computation of the evolution of collision products shows that, VMSs formed through runaway collisions and tidal capture, are dominated by mass loss from stellar winds  that drastically reduce their final remnant mass \citep{Glebbeek2009}.

Despite using a stellar evolution model where stellar winds strongly affects the most massive stars and direct collapse into IMBHs is not possible we can still form BHs more massive than $100$ \msun through BH-BH and BH-star collisions. We will show, in the next sections, that BH-VMS collisions provide the main channel for the formation of IMBHs. In other words, our results show that black holes with a mass above $100$ \msun form when a stellar black hole merges with a VMS in agreement with revious works \citep{Giersz2015, Mapelli2016}.  Runaway tidal captures \citep{Stone2017} can also produce IMBHs.




 \begin{table}
  \begin{tabular}{lccccc}
    \hline
  Model Name       & $R\mathrm{_h}\,$ & $\sigma$ & N & $t_{\mathrm{rh}}$ &$M_{f}/M_{i}$     \\
        & pc     & km/s   &     & Myr  & \\
  \hline
  R06W6        & $2.45$ & $6.09$ & $106800$    & $519$   & 0.695 \\ 
  R06W7        & $2.57$ & $5.93$ & $106400$    & $573$    & 0.693\\    
  R06W8        & $2.69$ & $5.71.$ & $106700$    & $620$   &  0.690\\
  R06W9        & $2.82$ & $5.56$ & $105900$    & $671$     &  0.688\\
  R1W7          & $2.92$ & $5.48$ & $107800$      & $772$   & 0.702  \\
  R1W8          & $3.11$ & $5.41$ & $107500$    & $820$     & 0.698\\
  R1W9          & $3.24$ & $5.33$ & $107200$    & $860$     & 0.692\\    
  R1W10        & $3.41$ & $5.25$ & $106700$  & $894$       & 0.689\\ 
  \end{tabular}
  \caption{ Half mass radius $R\mathrm{_h}\,$, dispersion velocity $\sigma$, half mass relaxation time  $t_{\mathrm{rh}}$,  number of particles N, and final to initial total cluster mass ratio $M_{f}/M_{i}$ after $300$ Myr; where $M_{i}$ is the initial mass of the cluster while $M_{f}$ is the final mass of the cluster.}
   \label{table:final}
\end{table}

\subsection{Collisions}

The outcome of a collision depends on the relative velocity, the relative sizes and the internal structure of the two colliding objects. In general, full 3D radiation magneto-hydrodynamical simulations are required to robustly determine the properties of the final object (such as mass, size, and internal structure). In the absence of results covering the full parameter space NBODY6++GPU adopts a simplified treatment. If a collision does not involve red giants the two objects are merged when the radii of two colliding objects overlap and the merger of the two masses is assumed to be instantaneous. If $M_{\mathrm{1}}$ and $M_{\mathrm{2}}$ are the masses of two stars or two compact objects (black holes, neutron stars, withe dwarf, etc.), the final object will have a mass $M_{\mathrm{f}}$ equal to:
\begin{equation} \label{eq:collision}
    M_{\mathrm{f}} = M_{\mathrm{1}} +  M_{\mathrm{2}}.
\end{equation}
Otherwise, if $M_{\mathrm{1}}$ is the mass of a compact object and $M_{\mathrm{2}}$ is the mass of a star, the final mass of the compact object, $M_{\mathrm{f}}$, is given by:
\begin{equation} \label{eq:f_c}
    M_{\mathrm{f}} = M_{\mathrm{1}} +  f_{\mathrm{c}} \times M_{\mathrm{2}},
\end{equation}
where $0<$ \f $<1$ represents the amount of stellar material falling back and accreted onto the black hole during the coalescence. The value of \f has only been investigated in the simulation literature for some specific cases \citep{Guillochon2019, Lixin2018, Metzger2016, Hotaka2015}. In the absence of solid results for all mass ratios we treat \f as a free parameter and explore scenarios with \f $=0.1, 0.5$ and $1.0$ (see Fig. \ref{fig:vms_bh_f}).

Collisions involving a giant or giant-like star with a dense core and a large envelope are assumed to lead to common-envelope evolution. In this situation the envelope of the red giant wraps around both cores of the two objects \footnote{For main sequence stars and black holes, the core mass coincides with the mass of the objects itself.}. The outcome depends on the orbital energy of the two cores as well as the binding energy of the envelope. If the latter is too low, the wrapper might be removed from the system before the two cores merge, leaving behind a binary composed of the appropriate remnants. Otherwise, the two cores are destined to spiral into each other. The final mass depends on their relative density. If they are of different compactness, equation \ref{eq:f_c} applies, otherwise the code uses \ref{eq:collision}.  

When two stars of similar compactness merge, we assume that they coalesce and mix completely. As a consequence the final star can be rejuvenated since the core of the final object absorbs new fuel. This phenomenon is well established as it explains the peculiar evolution of blue stragglers as the product of binary evolution and direct stellar collisions \citep{Smith2015, Davies2004, Davies2015}. This rejuvenation procedure also applies to MS stars, even if they are assumed to have no core. For more details see section 2.6.6 of \citet{hurley2002}.

When two black holes (or neutron stars) form hard binaries, gravitational waves emission is computed with the Peters \& Mathews formulae \citep{Peters1963}. The semi-major axes and the eccentricities are then changed according to the amount of gravitational wave energy emitted. 

\section{Initial Conditions}

\begin{figure}
    \includegraphics[width=\columnwidth]{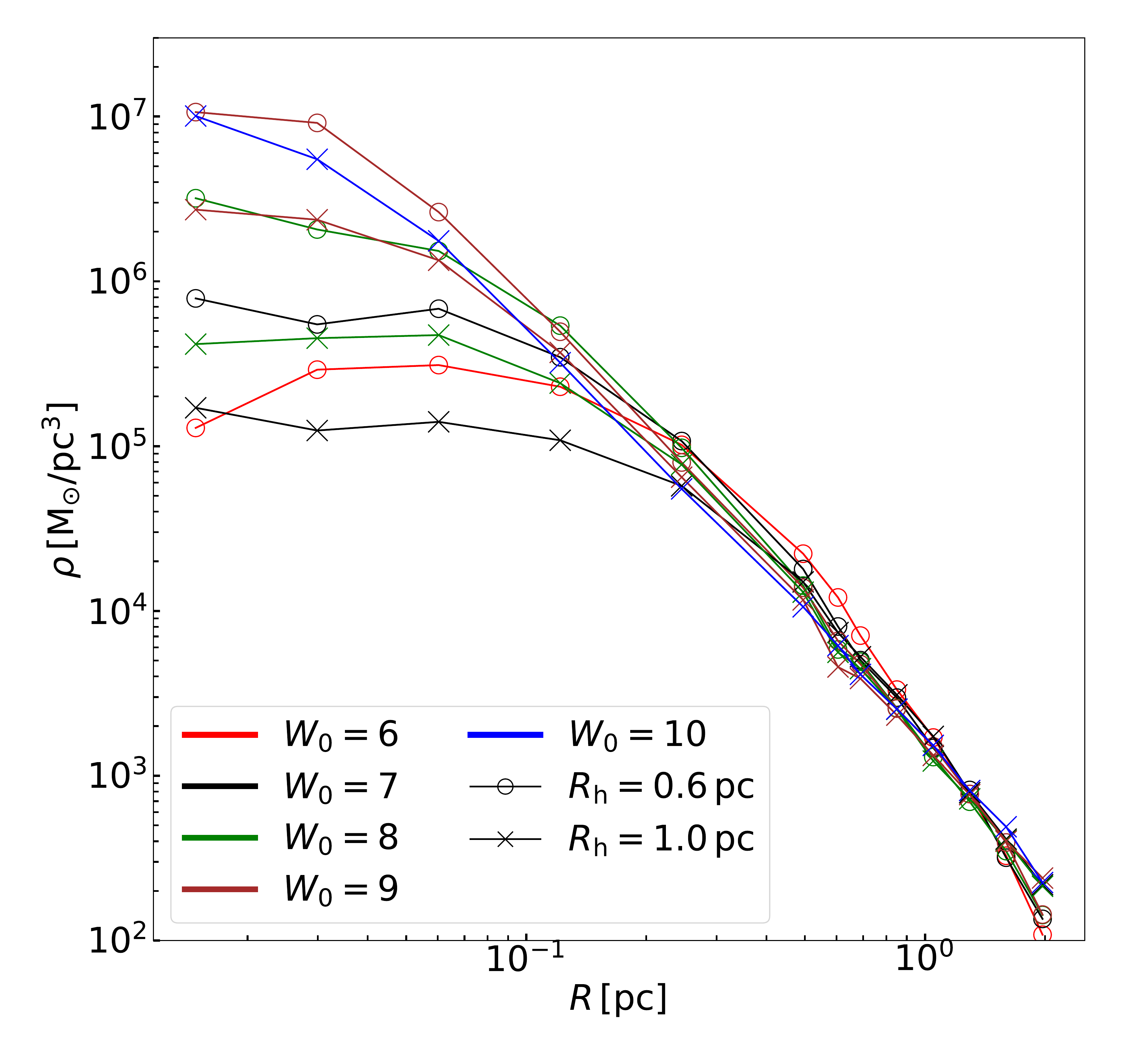}
    \caption{Initial density distributions for the cluster simulations presented in Tab. \ref{table:initial}, initialized with a King density profile \citep{King1966} with two half-mass radii and different central potential parameters \W (different colours). Systems with \Rha are represented with filled circles while models with \Rhb are represented by crosses.}
        \label{fig:density}
\end{figure}

\begin{figure*}
    \includegraphics[width=\textwidth]{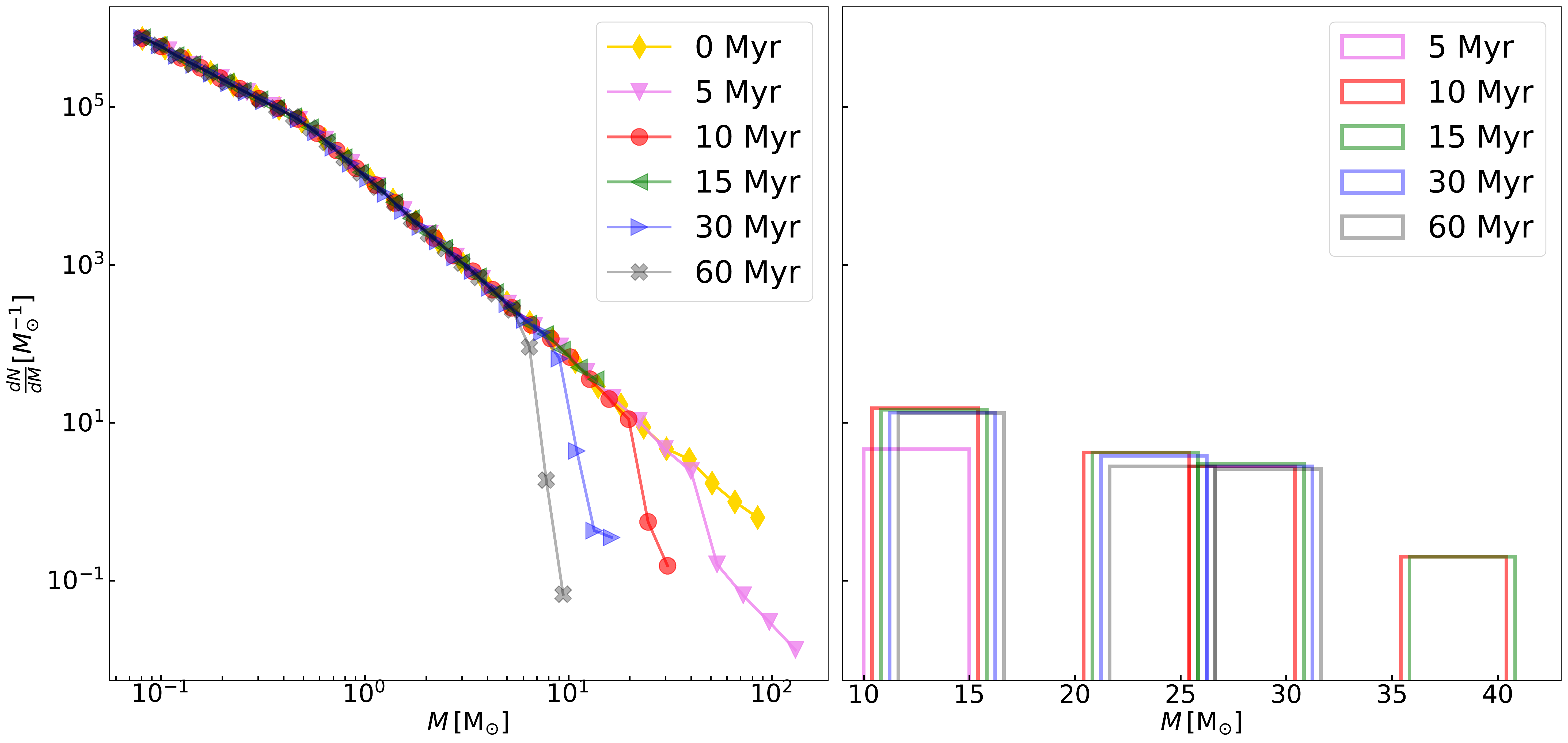}
    \caption{Left panel: stellar mass function (the mass function is computed including all stellar types apart from BHs, neutron stars and withe dwarfs) at $0$, $5$, $10$, $15$, $30$ and $60$ Myr. Right panel: compact objects mass function (the mass function includes only BHs, neutron stars and withe dwarfs) at $0.0$, $5$, $10$, $15$, $30$ and $60$ Myr.}
        \label{fig:IMF}
\end{figure*}

\begin{figure*}
    \includegraphics[width=\textwidth]{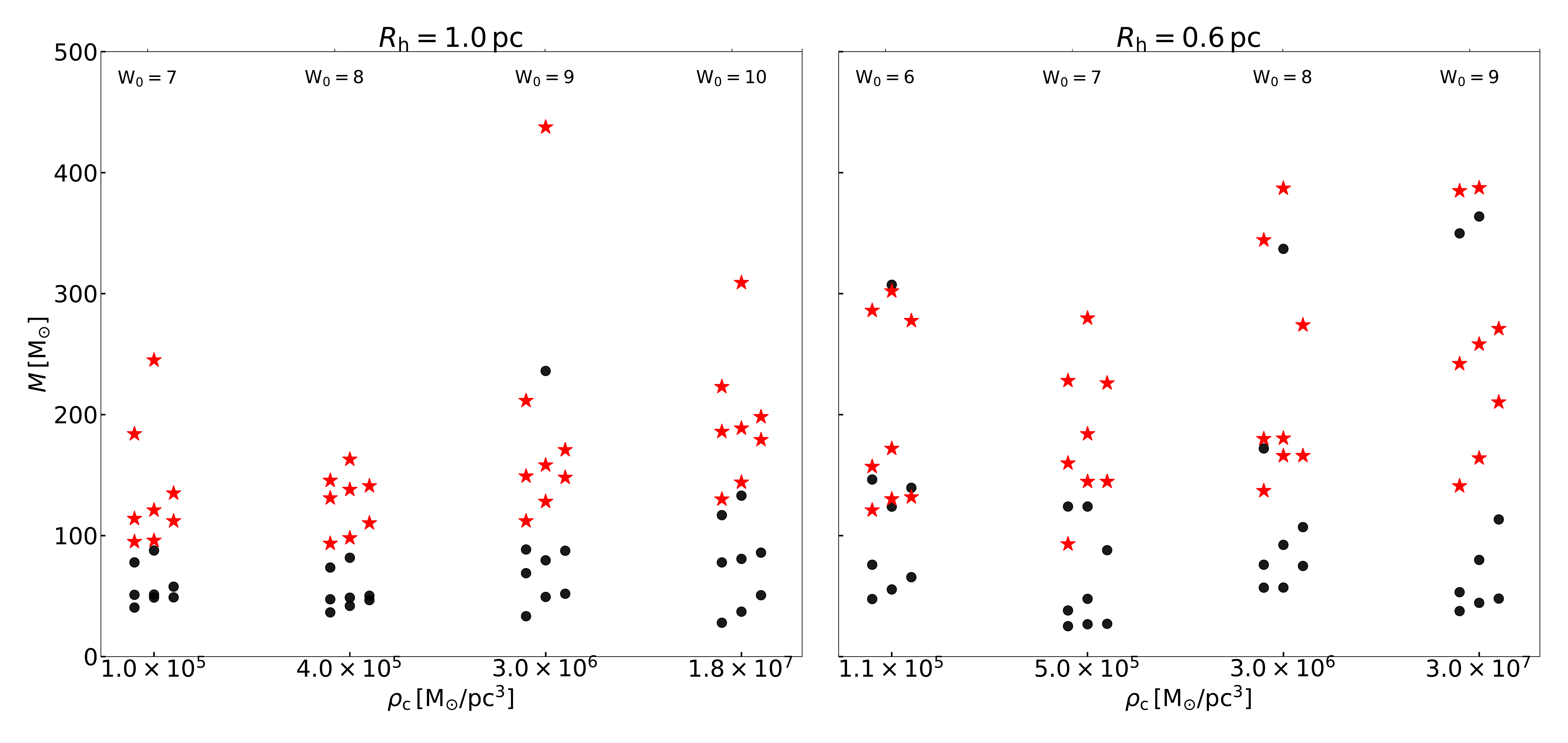}
    \caption{Peak masses of the very massive stars (VMS) formed by collisions (red stars) and black holes (black circles) for all simulations with $f_c = 1$ sorted by initial core density (see Tab. \ref{table:initial}). The left (right) panel shows models with a half-mass radius of $r_{\mathrm h} = 1 $  pc ( $r_{\mathrm h} = 0.6 $ pc). Each model has $8$ realizations (plotted with random offset). \W does not seem to have a strong impact on the final black hole mass. A larger cluster half-mass radius apparently makes black holes mass growth less likely (left panel). }
        \label{fig:vms_mbh}
\end{figure*}

We created initial conditions for 10 isolated   \footnote{According to the code, a star escapes from an isolated cluster when its energy is larger than zero, $E>0$ and its distance from the center is larger than $30R_{\mathrm{h}}$.} star cluster models following a King density profile \citep{King1966} with two different half-mass radii and five different central densities varying the dimensionless central potential parameter $6\leq$  \W $ \leq10$.  For a fixed value of the half-mass radius,  the central density increases with increasing \W\footnote{The exact definition of the parameter \W$=\frac{\psi(0)}{\sigma^2}$  where $\psi(0)$ is the potential at the center of the cluster and $\sigma^2$is a parameter connect to the velocity dispersion defined from the distribution function; see \citep{King1966} section 3 and \citep{HeggieHut2003} chapter 8 for more details.}  (see Fig. \ref{fig:density}). 
All simulated clusters were initialized with a very low metallicity of $Z=0.0002$  and $N = 1.1 \times 10^5$ stars sampled from a Kroupa IMF \citep{kroupa2001} as zero-age main-sequence stars in a mass range of $0.08$  \msun to $100$ \msun (see fig. \ref{fig:IMF}). No primordial mass segregation was included. We assumed a primordial binary fraction of $10\%$ ($10.000$ binaries) with a uniform semi-major axis distribution on a logarithmic scale from $0.001$ AU to $100$ AU \footnote{To generate the semi-major axis distribution we impose the further constraint that the initial semi-major axes of each binary are larger than the sum of the diameters of the two stars.}, a uniform distribution of mass ratios, and a thermal distribution of eccentricities; With this binary distribution about $30\%$ of the primordial binaries are weak and they dissolve in the cluster at the beginning of the simulation. For each initial condition parameter set we created 8 realisations with different random number seeds. In Tab. \ref{table:initial} we list the initial conditions parameters of the $10$  models including the number of realizations that lead to the formation of IMBHs and their respective masses. All simulations were run for more than 300 Myr up to 500 Myr\footnote{We evolved for longer time the clusters that formed an IMBH to check whether the BH would keep growing in mass or it would be ejected from the cluster through a strong interaction.}.  

The clusters studied in this work are initialized with a low metallicity value  (about two orders of magnitude lower then the solar metallicity), they are rather compact with high initial central densities and small half-mass radii. For this reason, they do not resemble the typical properties of young massive star clusters (YMSCs) in the observable range. The latter have higher, and closer to solar, metallicity and are typically less compact; observed YMSCs with masses similar to our models have virial radii typically in the range from 3 to 30 pc \citep{Zwart2010}.  Even considering the expansion driven by stellar and dynamical evolution  our less compact systems would have a virial radius from 2 to 30 times smaller then most of the observed massive clusters.

However, it is still possible to find a few  compact YMSCs in the local Universe.
An example is  Westerlund 1, which has a virial radius of $\sim 1.7$ pc ( our \Rha model at the same age has a virial radius of about 1.15 pc )   and a mass $\sim 6 \times 10^4$  \msun \citep{Mengel2007, Zwart2010}. Another exception is  MGG-11 in M82. This cluster has a half-light radius of about $1.2$ pc and a mass of about $3.50 \times 10^5 $ \msun \citep{McCrady2003}.

Our models start with velocity dispersions  $\sigma$ between  $15$ km/s and  $12$ km/s (see Tab. \ref{table:initial}) which drop to values between $6.1$ km/s and $5.3$ km/s after $300$ Myr  as shown in Tab. \ref{table:final}.  We also show the half mass radius, the relaxation time, the number of particles and fraction of total mass left in the cluster after $300$ Myr.
It is interesting to notice that R06W9, which formed the most massive IMBHs, after $300$ Myr has a dispersion velocity and half mass radius very similar to the least compact model  R1W7 (see table \ref{table:final}), which never formed a BH more massive than $100$ \msun. 
The velocity dispersions are shown in Tab. \ref{table:final} are approximately in the same range of the observed values of the globular clusters in our Galaxy \citep{Baumgardt2018}. However, our clusters are too small to resemble the initial conditions of present-day globular clusters and estimating the particle number of our clusters at late times  using equation 22 in  \cite{Gieles2011}, our systems might not survive for $10$ Gyr even if we assume no external tidal forces.

We have chosen these initial conditions mainly to investigate the formations of  IMBHs through dynamical interactions.
Nevertheless, our theoretical models might approximate the properties of clusters formed at a high redshift, which may be located around any galaxy in the LIGO/Virgo sensitivity volume ($ \sim 1$ Gpc$^3$).


\section{Results}



The dynamical evolution of a star cluster is dominated by two-body relaxation on time scales longer than the relaxation time \citep{Spitzer1987} :
\begin{equation}\label{eq:trh}
t_{\mathrm{rh}} = \frac{0.138 N^{1/2}}{\ln{\Lambda}} \left(\frac{R_{\mathrm{h}}^3}{G\Bar{m}}\right)^{1/2}.
\end{equation}
Here $N$ is the number of stars in the cluster,  $R_{\mathrm{h}}$ is the half mass radius and, $\Bar{m}$ is the average star mass of the cluster  and  the argument of the Coulomb logarithm is $\Lambda=\gamma N$. Numerical experiments indicates a value for the parameter $\gamma = 0.11$ for single-mass systems \citep{Giersz1994}  and $\gamma = 0.02$ for multi-mass stellar systems \citep{Giersz1996}.
We have used Eq. \ref{eq:trh}  to estimate half-mass relaxation times of $56$ Myr and $120$ Myr for systems with $0.6$ pc and $1$ pc  half-mass radii, respectively (see Tab. \ref{table:initial}).

The internal structure  of star clusters  evolves into a dense hot core and an extended halo. Since bound self-gravitating systems have negative heat capacity  \citep{Lynden-Bell1968, Lynden-Bell1999}, the center will keep releasing energy to the  outer part and contracts in a core collapse.  In an isolated  equal-mass system,  this happens on the order of $15 t_{\mathrm{rh}}$ \citep{Cohn1980}.
In clusters with a broad mass spectrum, low mass stars gain kinetic energy when interacting with massive stars; the former tend to expand their orbits, while the latter tend to lose kinetic energy and segregate to the central part of the cluster.
In this case, core collapse is driven by the amassing of heavy stars in the core that typically occur in a time scale of the order of the segregation time \citep{ Spitzer1971, Zwart2004}:
\begin{equation} \label{eq:ts}
 t_{\mathrm{s}} = \frac{ \Bar{m} }{ M_{\mathrm{max}} } \frac{0.138 N}{\ln{(0.11 M/M_{\mathrm{max} }) } } \left(\frac{R_{\mathrm{h}}^3}{GM}\right)^{1/2}
\end{equation}
where $M_{\mathrm{max}}$ is the mass of the most massive object in the cluster and $M$ is the total mass of the cluster . Consequently  multi-mass clusters undergo core collapse in much shorter time then single-mass systems.  According to equation \ref{eq:ts} our \Rha and \Rhb models are expected to experience mass segregation at  $t_{\mathrm{s}} \sim 1.4$ and 3 Myr respectively (see Tab. \ref{table:initial}). 

Core collapse lead to dramatic growth in the central density which in turn triggers  violent  few-body interactions between single stars and binaries, either primordial or dynamically formed. By means of this interaction binary stars release energy in the core and balance the loss of energy from the centre preventing the core to collapse further \citep[see][and references therein]{HeggieHut2003}.

 We show in Fig. \ref{fig:vms_mbh} that the most massive stars formed in each simulation (red crosses) consistently have masses much higher than the initial limit of $100$ \msun. These system have formed by mergers of lower mass main sequence stars.  As described in section \ref{sec:stellarEvolution} our stellar evolution recipes do not allow for the formation of a stellar black hole more massive than 30 \msun. Nevertheless, Fig. \ref{fig:vms_mbh} demonstrates that several cluster systems generate BHs more massive than $100$ \msun. In our framework, the only possible way to grow such a massive object is through dynamical interactions (see Sec. \ref{subsec:mbhf}).

The likelihood to form a very massive star by mergers or tidal captures seems to be correlated with the compactness of the cluster. Only $6$ star clusters with \Rhb form stars with masses above $200$ \msun, while clusters with \Rha form $15$ stars more massive than $200$ \msun (see Fig. \ref{fig:vms_mbh}). Those stars play an important role in the production of IMBHs as their collisions with stellar BH are the main IMBH formation channel. Such a formation path for IMBHs has been predicted by Monte-Carlo models \citep{Giersz2015} and is discussed in more detail below.




\subsection{Intermediate Mass Black Hole Formation}
\label{subsec:mbhf}

\begin{figure*}
	\includegraphics[width=\textwidth]{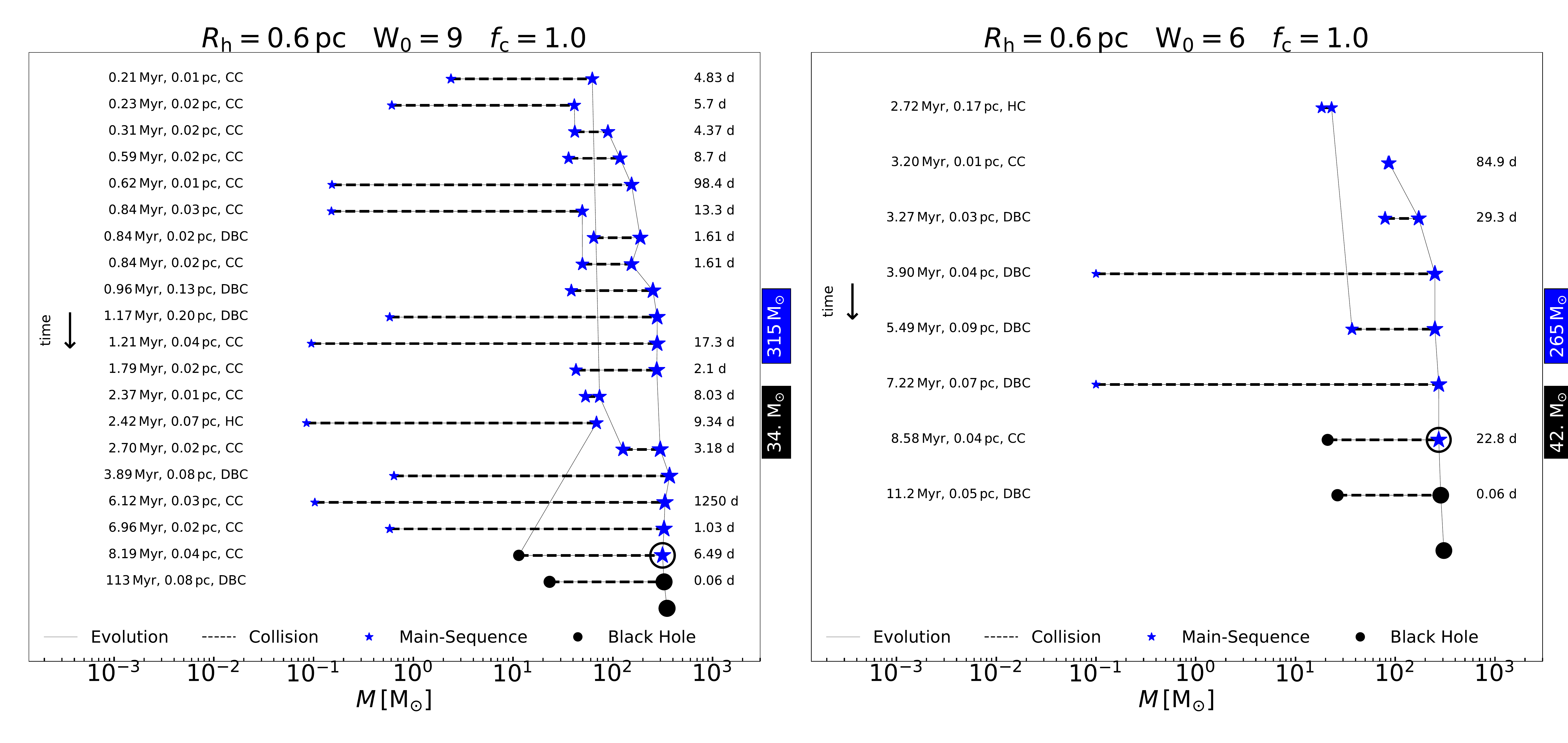}
    \caption{The left panel shows - from top to bottom - the formation history of an IMBH in a simulation with \W $=9$,  \Rha  and \fc. A massive main-sequence star (MS, blue) grows by mergers (dashed horizontal lines) with other MS stars and evolves into  $\approx 300$ \msun  MS star colliding with a  $10$ \msun BH, forming an IMBH about 8.2 Myr after the start of the simulation (the open black circle indicates the collision that leads to the formation of the IMBH.) . At about $113$ Myr the IMBH collides with an other BH. In a similar fashion, an IMBH of about $300$ \msun is generated in a lower density cluster with \W $=6$,  \Rha  and \fc just with less merger of more massive stars (right panel). In each panel we indicate the time of the collision after the start of the simulation in Myr and the radius from the center in pc. We also characterise the type interaction in the following way: \textbf{BC}: Binary collision -  the two colliding objects formed a binary before the collision, \textbf{PBC}: the binary is primordial, \textbf{DBC} the binary has formed dynamically,   \textbf{HC}:  hyperbolic collision,  \textbf{CC}: chain collision - all collisions not classified as \textbf{DBC} or \textbf{HC}. If the two colliding objects formed a binary before coalescence the period is given in d. In the blue (black) boxes we indicate the  amount of stellar (black hole) material accreted by the IMBH.
Here it is important to mention that the figure shows all collisions experienced by each participant; if an object appears in the plot only once it implies that the object never experiences any collision before. 
The are no BH - BH mergers before the formation of the IMBH. Therefore, even if we did not include the gravitational kicks, the result does not change.}
  \label{fig:CollisionTree_IMBH}
\end{figure*}

\begin{figure}
	\includegraphics[width=\columnwidth]{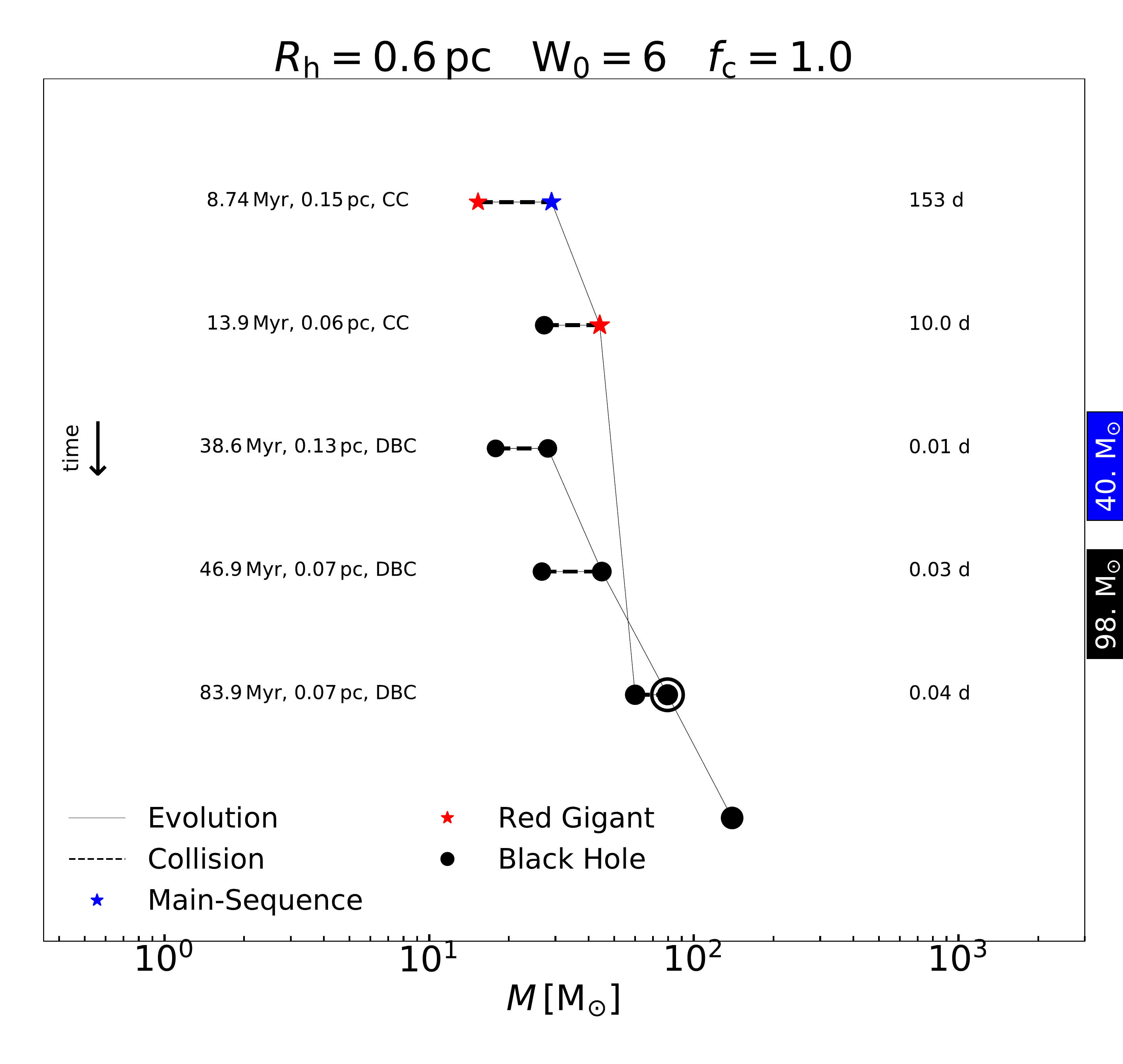}
    \caption{Formation and evolution of an IMBH in a simulation with \W $=6$,  \Rha  and \fa, similar to Fig. \ref{fig:CollisionTree_IMBH}. However, here the black hole reaches a mass of about $140$ \msun mostly through collisions (horizontal dashed lines) with other BHs. This is the only case, out of $80$ simulations analysed, where a BH above $100$ \msun built up its mass mostly through collisions with other BHs.}
    \label{fig:CollisionTree_BH-BH}
\end{figure}

\begin{figure*}
	\includegraphics[width=\textwidth]{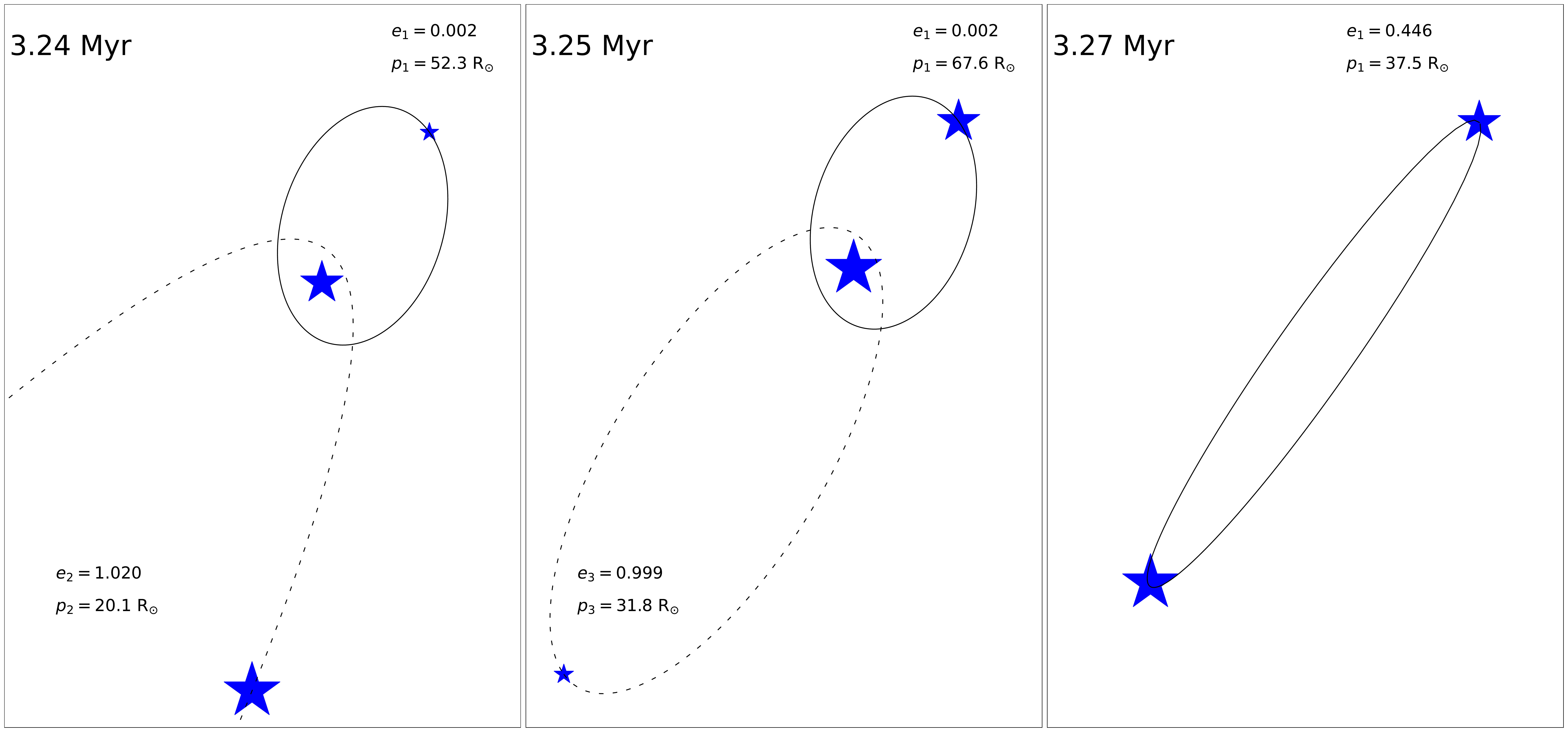}
    \caption{Sketch of the triple interaction that induces the collision of two massive stars of 170 and 80 \msun. Left panel: the 170 \msun approaches a binary in a hyperbolic orbit. Central panel: the exchange between the lightest component of the binary and the intruder of 170 \msun.  Right panel:  the perturbed 170-80 \msun binary increase its eccentricity, the two components are about to merge.}
  \label{fig:star_star_interactions}
\end{figure*}

\begin{figure*}
	\includegraphics[width=\textwidth]{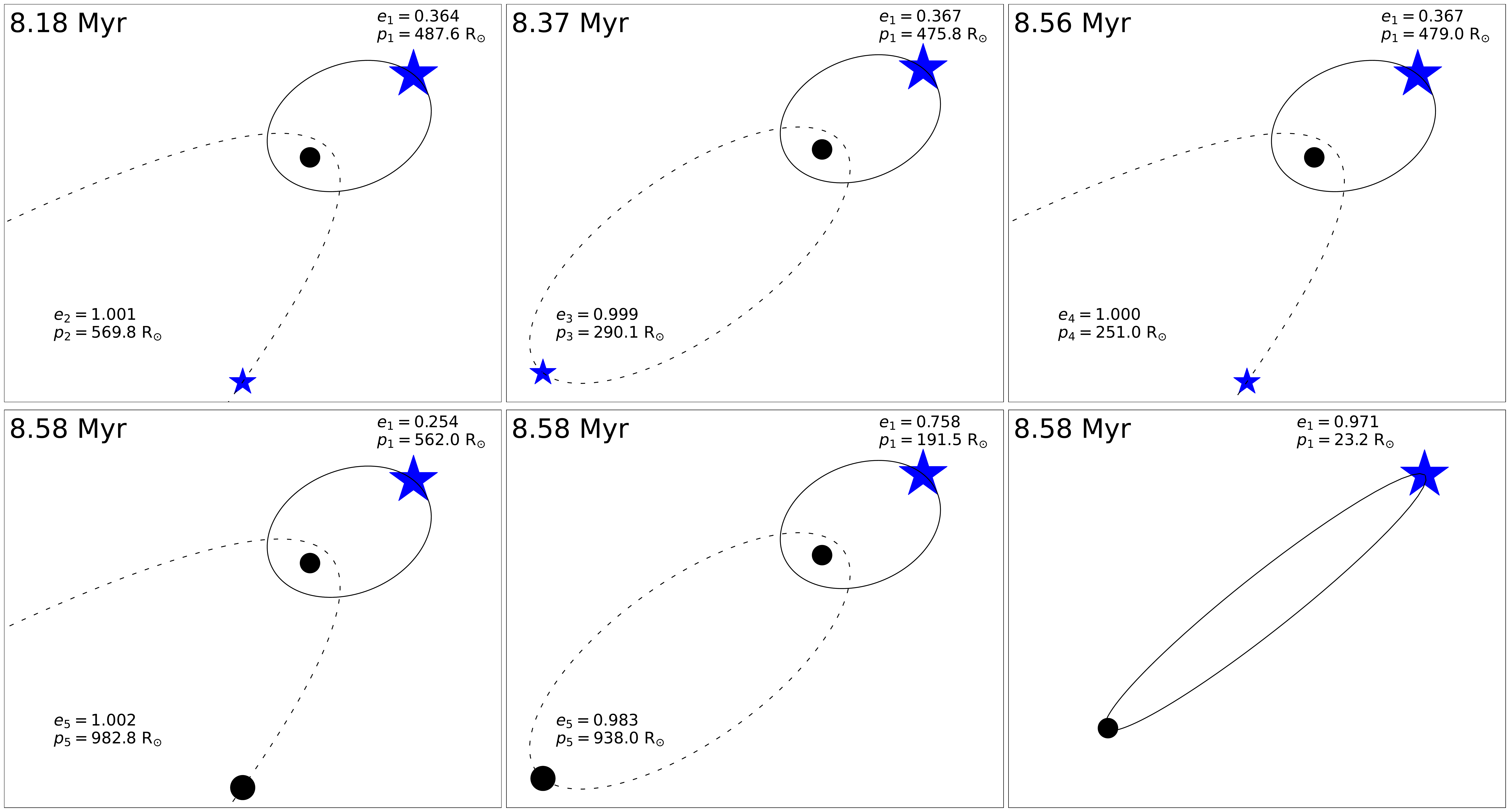}
    \caption{Sketch of the last five triple interactions experienced by a BH-VMS binary before the collision. In the last interactions, the BH-VMS binary forms a hierarchical triple with a stellar BH. Consequently the binary increases its eccentricity to a value close to unity that leads the BH to merge with the VMS.}
  \label{fig:BH_star_interactions}
\end{figure*}

\begin{figure}
	\includegraphics[width=\columnwidth]{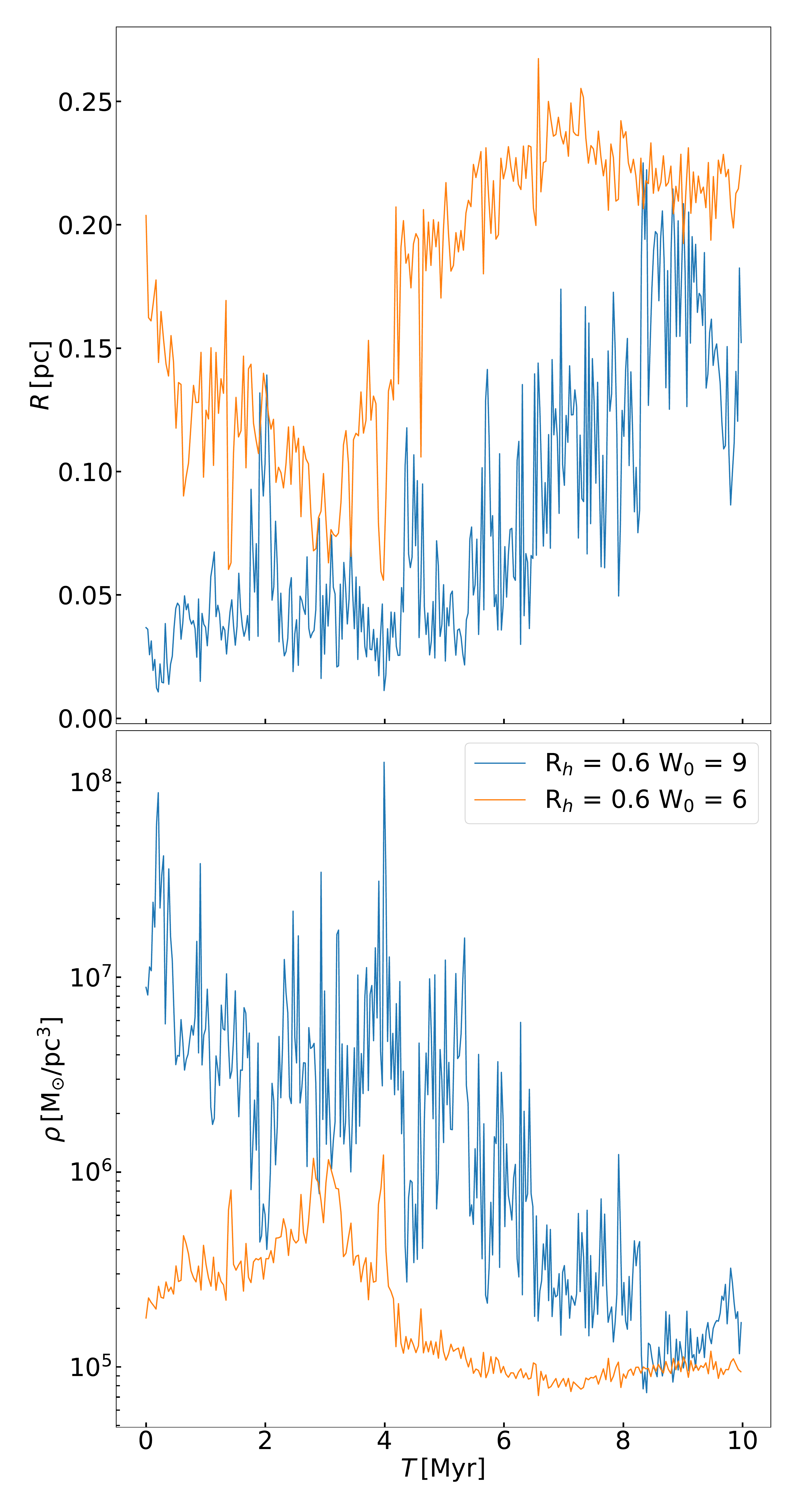}
    \caption{Time evolution of core radius (top panel) and core density (bottom panel) for two models with the same initial size \Rha but different initial central potential parameter \W (blue: R06W9, orange: R06W6). The figure shows a constant rapid expansion for the R06W9 model as opposed to an initial contraction for R06W6 simulation. The initial central density difference in the two models decreases in the first few Myr.} 
    \label{fig:core_density}
\end{figure}


Figure \ref{fig:CollisionTree_IMBH} illustrates how IMBHs of a few hundred solar masses are typically generated in our simulations. The formation consists of three main steps. First, a sequence of binary stellar collisions, triggered by triple interactions and hyperbolic collisions generates a VMS which can live up to $10$ Myr due to mixing rejuvenation (see section  \ref{sec:stellarEvolution}). Second, in a merger with a stellar mass BH, a great part of the mass of the VMS is absorbed by the BH. In a third step, the IMBH can grow in mass by collisions with other stellar BHs. Our results indicate that, for our models, the dominant process for the formation of IMBHs is a collision between a VMS  and a stellar mass BH. This type of collisions can lead to the formation of IMBHs of up to  $350$ \msun within the first $10$ Myr of cluster evolution. Our simulations also indicate that after all massive stars disappear from the cluster, the IMBH can still grow moderately in mass by merging with other stellar mass black holes and other types of stars.

Our most compact models register about 300 collisions within 300 Myr, $40\%$ of which happen in the first 15 Myr. 
Most of these collisions are triggered by three-body scattering events \footnote{About $2\%$ of collisions are hyperbolic.} between a hard binary and a third particle. These interactions have the overall effect of increasing the binding energy of the binaries and they can also raise their eccentricity \citep{Heggie1975, Nash1978, Hut1983, McMillan1996}.
In general, when the distance of closest approach of the third object is comparable with 
the semi-major axis of the hard binary, the net effect of the interaction is to harden the binary \citep{Heggie1975, Nash1978, Hut1983, McMillan1996}.
On the other hand, when the pericenter of the intruder is considerably larger then the size of the binary, the interaction is approximately adiabatic, and there is no exchange of energy between the binary and the intruder. However, the binary and the third object can form a hierarchical triple, which excites the eccentricity of the inner binary to values close to unity \citep{Lidov1962, Kozai1962} and induces the two components of the binary to merge.
Figure \ref{fig:star_star_interactions} illustrates the dynamical process leading to the collsion of two massive stars with $170$ and $80$ \msun, respectively. The $170$ \msun star approaches a binary ($0.1-80$ \msun) in a hyperbolic orbit (left panel); during the interaction, the intruder forms a hard binary with the $80$ \msun star, while the lightest component escapes in a wider orbit (panel 2). Due to the perturbation of the third object the eccentricity of the new $170-80$ \msun  binary increases from 0.002 to 0.45 and the two massive stars crash into each other.    
Figure  \ref{fig:star_star_interactions} also shows that the colliding objects do not necessarily need to be in a primordial binary. Exchanges are very frequent in triple interactions. Typically, encounters leading to exchange increase the mass of the components of a hard binary because the lowest mass is most likely to escape \citep[see][and references therein]{HeggieHut2003}.

In summary, when a single object interacts with hard binaries, the binaries tend to harden and become more massive, and they also gain angular momentum and eccentricity. 
For all these reasons triple interactions drive the chain of star-star collisions leading to the formation of VMSs. They are also the main process that triggering BHs-VMSs mergers (see Fig. \ref{fig:BH_star_interactions}).

\begin{figure}
	\includegraphics[width=0.95\columnwidth]
    {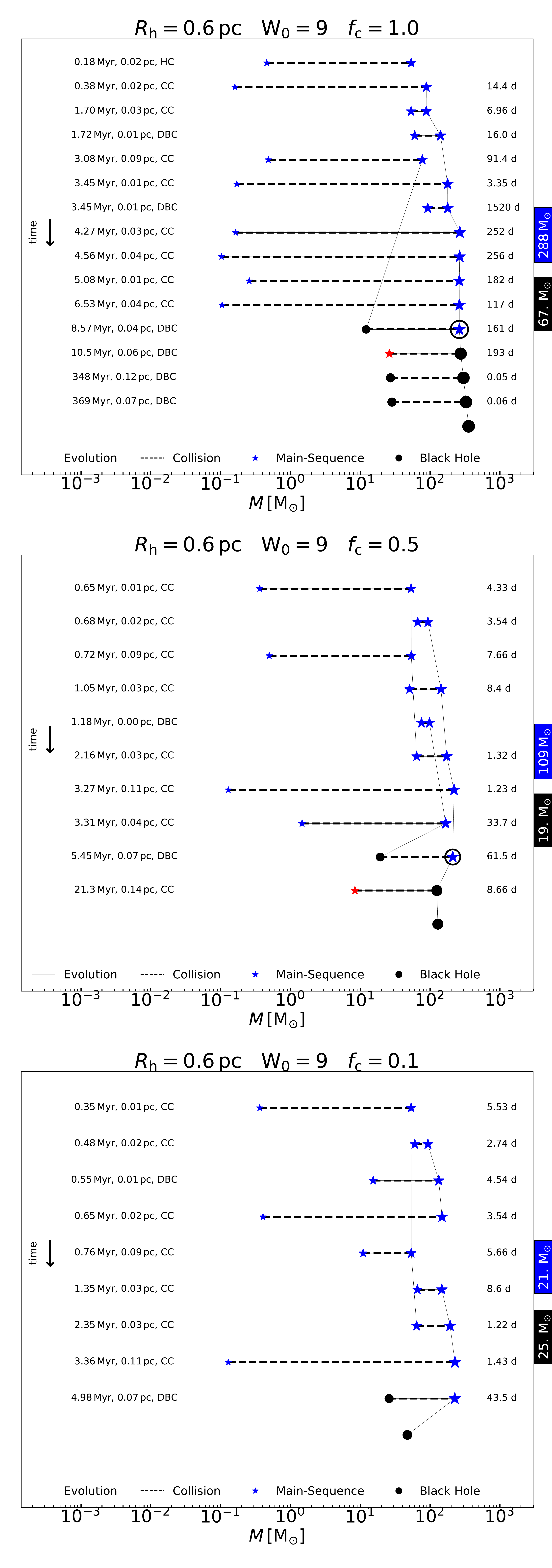}
    \caption{Comparison of three models with identical initial conditions and decreasing \f of 1, 0.5, and 0.1 from top to bottom. For \f =0.1 no IMBH forms. The time of IMBH formation is indicated by a black circle in the two top panels. Here we show a different realization of model R06W9 than in Fig. \ref{fig:CollisionTree_IMBH}.}
    \label{fig:ManyCollisionTree}
\end{figure}

Here it is important to mention that gravitational kicks are not implemented in our simulations. Theoretically,  an IMBH could be ejected from the cluster by gravitational wave recoil after a collision with one of the stellar mass  BHs left in the system. However,  except for of the simulation illustrated in Fig.  \ref{fig:CollisionTree_BH-BH},  we expect the clusters to have good chances of retaining their IMBHs as the IMBH - BH mass ratio is large. Therefore the gravitational velocity kick is likely to be smaller than the local escape velocity according to \citet{Campanelli2007, Baker2008, Kulier2015, Morawski2018, Zivancev2020}. 
The absence of gravitational recoil velocity does not influence the series of collisions and interactions that lead to the VMS-BH mergers. The BHs involved in these collisions are the remnant of massive stars and they did not merge with any other BH. This can be clearly seen from fig. \ref{fig:CollisionTree_IMBH} and fig. \ref{fig:ManyCollisionTree}. 
Both figures report all the collisions experienced by each object that participates in the collision chain. Therefore if a BH appears in the plot only once it implies that the object never experiences any collision before.

The IMBHs in our simulations are mainly growing through two channels:
\begin{itemize}
\item Accretion of stellar material through binary collisions with stars and, 
to a lesser extent, through hyperbolic collisions with stars.  
\item Collision with low mass BHs.
\end{itemize}

BHs can also become more massive through multiple mass transfer events that do not lead to coalescence. However, the mass absorbed in those events is negligible. Moreover, BHs could also grow through tidal captures. However, according to our simulations, tidal captures do not seem to be relevant for the IMBHs final mass. Tidal captures are very rare (10-15 events per run) and only a small fractions of these events lead to collisions.   


Collisions with stellar BHs typically add little ($\sim 10 - 30 \%$, see Tab. \ref{table:initial}) to the total IMBH mass. BH-BH collisions are not very likely to occur. This type of binary requires very small semi-major axes, or very high eccentricities to enter in the post-Newtonian regime, experience gravitational wave energy loss, and eventually merge. This requires the BH binary to undergo several strong triple interactions (or binary-binary interactions). For this reason, as we can see in Fig. \ref{fig:CollisionTree_IMBH}, and later also in Fig. \ref{fig:ManyCollisionTree}, that most of the material accreted onto a massive BH originates from stars. We form in total $17$ BHs with a masses above $100$ \msun (see Tab. \ref{table:initial}). Despite being unlikely, one of the IMBHs grows its mass almost entirely by swallowing other black holes (as shown in Fig. \ref{fig:CollisionTree_BH-BH}). The formation process  occurred in about $\sim 90$ Myr and it involves a chain of low mass BH mergers with masses of $17:28,\,25:45,\,68:70\,$ \msun.

According to Fig. \ref{fig:vms_mbh} the less concentrated model R06W6 and the more dense model R06W9 have comparable probability to produce IMBHs and VMSs despite the marked difference in the initial central density.
Fig. \ref{fig:core_density} displays the time evolution of the core radius (plot on the top) and the core density (plot on the button) for two simulations  with \W $=9$ and \W $=6$. The two systems experience different initial evolution: the core of the cluster with higher central density expand from the beginning of the simulation due to primordial binaries interactions, contrarily clusters with \W $=6$ undergo core collapse.  (this is also confirmed by the evolution of the Lagrangian radii shown in Fig. \ref{fig:Lagrangian} ). Consequently, the initial concentration discrepancy between the two models rapidly decreases. This fact might explain why clusters with \W $=6$ and \W$=9$ lead to comparable results although the initial central densities of the two models differ of  more then two orders of magnitude.

\begin{figure}
    \includegraphics[width=\columnwidth]{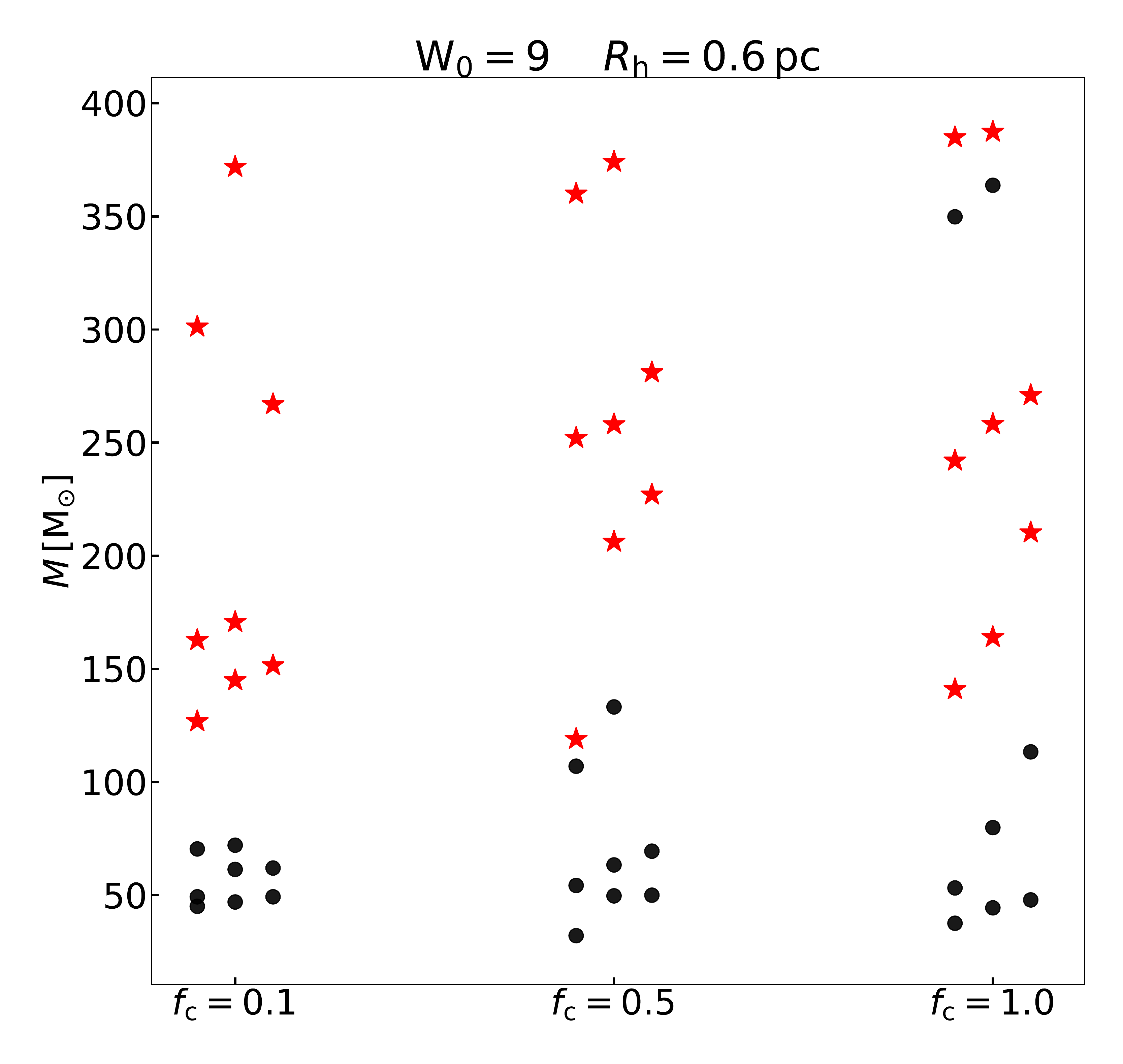}
    \caption{
    Comparison between R06W9, R06W9F05, and R06W9F01.  Each black circle (red star) shows the mass of the most massive BH (MS star) formed in a single simulation.
 }
    \label{fig:vms_bh_f}
\end{figure}

\subsection{Collision Fraction \f}

So far we have assumed an accretion fraction of \f =1 i.e. in star - BH collisions all stellar material is accreted onto the BH. Under these circumstances IMBHs regularly form in the simulations. However, the accretion fraction is highly uncertain and we have simulated R06W9 with lower fractions of \f =0.5 and \f=0.1.  The results indicate that the accretion fraction can significantly change IMBH growth (see Fig. \ref{fig:vms_bh_f}). In fact, while the stars still grow in mass by collisions, no IMBH is formed in the simulation with \fa (R06W9F01, the most massive BH has a mass of about $60$ \msun). Even for simulations with \fb just two (out of 8 simulations) IMBHs form with masses only slightly above a hundred solar masses. Apart from the obvious effect that the BHs grow less in a collision with a VMS, there are also secondary effects. Less massive BHs  have a higher probability to escape from the cluster after strong interactions and they have lower probabilities of experiencing additional collisions due to lower gravitational cross-sections decreasing the probability to experience close encounters. 

\begin{figure}
    \includegraphics[width=\columnwidth]
    {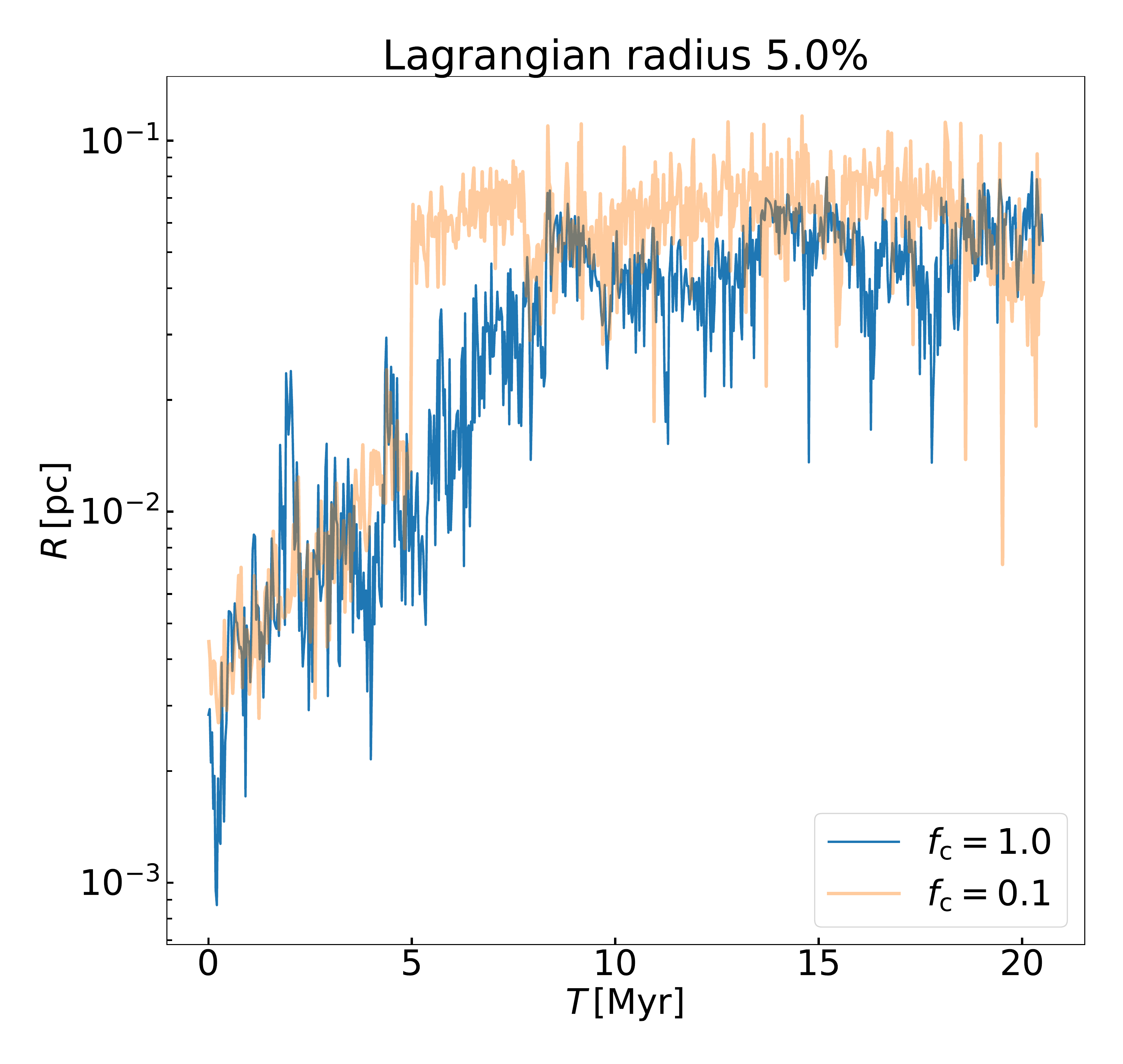}
    \caption{Comparison between $5\%$ Lagrangian radii  (see subsection \ref{sub:Cluster Evolution} for the definition of Lagrangian radii)  for a simulation with \fc (blue) and  a simulation with \fa (orange). Observing the orange line we can notice a fast expansion at about $5$ Myr triggered by mass-loss.  This dilatation is generated by a BH-VMS collision event that occurred at $4.98$  Myr. During the collision $90$ \% of the total mass of the VMS is assumed to be ejected instantly from the cluster.  For simulations with \fc, these type of expansions do not occur because the mass of stars is fully retained in the cluster when BH-star collisions occur.}
    \label{fig:Lagr5}
\end{figure}

Secondly, BH-VMS collisions with lower accretion fractions result in instantaneous mass removal from the inner part of the cluster  \footnote{In all our simulations BH-VMS collisions occur in the central region of the cluster.} leading to cluster expansion, lower densities, and a lower merger rate. As we can see in Fig. \ref{fig:Lagr5}, this effect is particularly enhanced in \fa simulations because in this case when a star collide with a BH $90 \%$ of the mass of the star is instantly ejected from the cluster.


\subsection{Cluster Evolution}
\label{sub:Cluster Evolution}

\begin{figure*}
    \includegraphics[width=\textwidth]{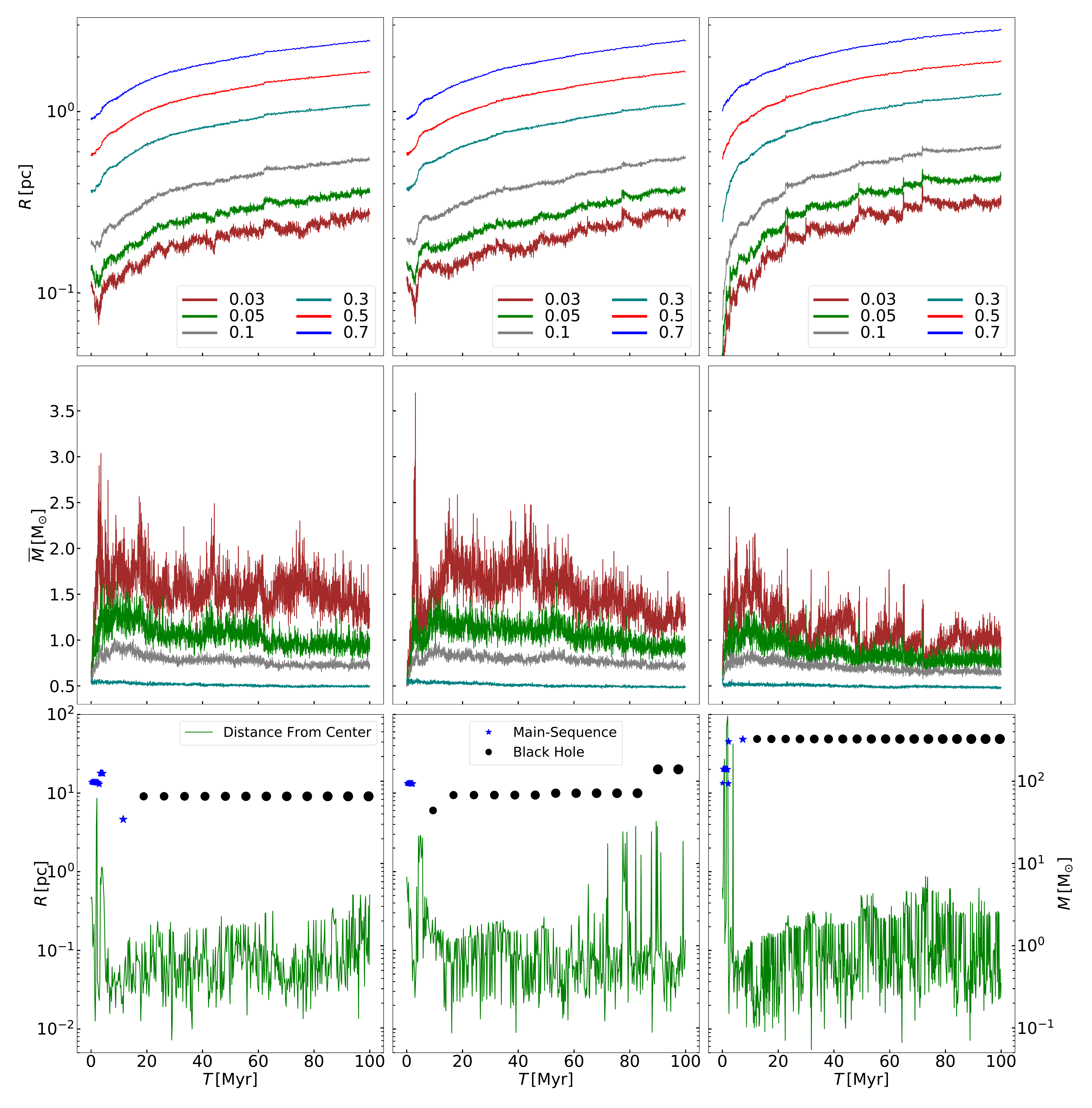}
    \caption{Lagrangian radii evolution (upper plots), particle  mass average evolution (central plots),  distance and mass of the most massive object (button plots). The left central and right panels report the results of three different simulations. The simulation on the left is a realization of the R06W6  model  that does not form IMBHs (the most massive BH has a mass of 60 \msun).  The central panel refers to another realization of the R06W6  model that generated an IMBH  of $140$ \msun through a chain of BH-BH collisions.  The plots on the right report the outcome of a  R06W9  realization where an IMBH of  about $350$  \msun formed in a VMS-BH collision.}
    \label{fig:Lagrangian}
\end{figure*}



In this section, we highlight the evolution of three different simulations resulting in very different peak BH masses, which we label, for simplicity, with $S_{1}$, $S_{2}$ and $S_{3}$. $S_1$ and $S_{2}$ are two different realizations of the model R06W6. The latter creates an IMBH of $140$ \msun (the evolution path of this BH is shown in Fig. \ref{fig:CollisionTree_BH-BH}). The most massive BH in $S_1$ only reaches $60$ \msun. $S_{3}$ is one realization of the model  R06W9 generating an IMBH, with a mass of about $350$  \msun (see evolution path in Fig. \ref{fig:CollisionTree_IMBH}).
The left, center and right  panels in Fig. \ref{fig:Lagrangian} refer to $S_{1}$, $S_{2}$ and $S_{3}$. respectively. Each panel consist of three plots. 
The plots at the top display the time evolution of the radii enclosing 3\%, 5\%,  10\%,  30\%, 50\% and 70\%  of the cumulative stellar mass (Lagrangian radii).
The  middle plots illustrate the evolution of the average stellar mass within  3\%,  5\%,  10\% and 30\%  Lagrangian radii. 
The plots at the bottom show the mass (and the type) of the most massive object in each cluster as a function of time and its distance from the cluster center. 

The evaluation of these three systems highlights the complex interplay between stellar evolution and dynamical interactions. 
The former has a strong impact on the early phase, while the latter plays a major role in driving the long-term change. Stellar evolution  mass loss  triggers  a strong  expansion on early cluster evolution \citep{Applegate1986, Chernoff1987, Chernoff1990, Fukushige1995}
Our simulations confirm these results. The 30\%, 50\%, 70\%  Lagrangian radii indicate a strong monotonic cluster inflation in the first $\sim 10$  Myr as a consequence of mass loss due to supernovae explosion and stellar winds followed by a moderate expansion driven by primordial binaries and further mass loss due to stellar/binary evolution.

Since our three simulations are multi-mass systems they undergo mass segregation in about $\sim 1.4$ Myr (see table \ref{table:initial}).
As a consequence of that, the average stellar mass in the inner part of the three clusters strongly increases in the first $\sim 2$ Myr (middle panels of Fig. \ref{fig:Lagrangian}). $S_1$ and $S_2$ also register an initial core collapse connected with mass segregation (rapid decrease for  3\%,  5\%, and 10\%  Lagrangian radii) in agreement with previous numerical work \citep{Zwart2002,Zwart2004, Gurkan2004, McMillan2008}. On the other hand, 
$S_3$ does not show any indication of core collapse. The system is already so dense that binary interactions prevent the collapse and force the core to expand (the innermost Lagrangian radii are in constant expansion since the beginning of the simulation. See top-right plot of Fig. \ref{fig:Lagrangian}).

Subsequently,  the death of the most massive stars leads to a mass average drop in the inner part of the cluster for both $S_1$and $S_2$ (see red line in  central plots).
The average mass then continues to oscillate, but while in  $S_1$ it presents a steady decline, $S_2$  raises again due to a BH-BH merger after about $15$ Myr. After that, it declines at a constant rate due to stellar evolution mass loss.
The initial drop is less evident in $S_3$ also because the presence in the center of the IMBH\footnote{In $S_3$ the IMBH form very fast, at about $8$ Myr. On the other hand, the IMBH in $S_2$ is generated after $84$ Myr.} compensates for the absence of the massive stars.

Massive objects located in the core of the cluster experience frequent strong interactions. This effect is reflected in the strong oscillations of the mass average in the core (see central plots) as well as the strong variations in the position of the IMBHs (see green lines in the button plots). 
 In $S_3$ the IMBH oscillates in position between $0.01$ and $0.5$ pc with an average position of about $0.1$ pc. Similarly the BH in $S_1$ moves around $0.1$ pc with slighter stronger oscillations due to its lower mass. On the other hand, the BH in $S_2$ experiences strong radial change few millions years before its coalescence with an other BH at $84$ Myr. These heavy oscillations  are generated by the strong interactions that triggered the last BH-BH collision.








\subsection{Comparison with Observations}
Very massive stars, formed through collisions between lower mass stars, appear in almost all the realizations of our ten models. If the VMSs formation mechanism proposed in this and other works \citep{Zwart2002, Zwart2004,Gurkan2004,Mapelli2016,DiCarlo2020,Wang2020} is correct they might be observed inside or close to dense star clusters.
Early studies of the Arches cluster indicated an upper star mass limit of $150$ \msun \cite{Figer2005}. However, more recent observations indicate the existence of stars greatly exceeding this limit, suggesting the presence of VMSs up to $300$ \msun  in the vicinity of young massive star clusters \citep{Crowther2010}. Another study claimed the discovery of a VMS of initial mass in the range between $90-250$
\msun in the central cluster of the region W49 \citep{Wu2014}. These stars might be generated via runaway collisions as indicated by the outcome of our and previous studied discussed in section \ref{section:comparison_with_other_work}. However, gas accretion could be an equally valid mechanism for the formation of VMSs \citep{Krumholz2015}.

It has been shown that the massive BH that power the hyper-luminous source associated with MGG-11 might have a dynamical origin.
The high density and compactness of this cluster allow for the formation of a few thousand solar masses star via collisional runaway, which might directly collapse into an IMBH \citep{Zwart2004}. 
Also, our results suggest that IMBHs could be the origin of HLXs associated with dense star clusters.  It is not rare for our simulated IMBHs to accrete mass and merge with stars,  as shown in the top and central plots of Fig. \ref{fig:ManyCollisionTree}  (both plots show the IMBH merging with a red giant at 10 and 21 Myr). However, even if our results show that IMBHs do receive mass from other stars, the mass transfer events are very often interrupted by strong interactions and therefore they do not last more than 1 Myr.   In other words the IMBHs tend  to spend only a small fraction of their time accreting material from their companion.  This, in addition with the fact that  IMBHs have a non negligible probability to be ejected from the cluster after a BH-IMBH collision  \citep{Arca_Sedda2020, Mapelli2020},  might explain the absence of IMBHs accretion signature in may star clusters and most of globular clusters \citep{Wrobel2015, Wrobel2020}.

Our simulations reveal that IMBHs, right after formation, tend to bind with a low mass black hole in a BH-BH binary. The binary, located a the center of the cluster, experiences constant gravitational interaction with other objects, and it merges in an interval of time between $ \sim 10-100$ Myr generating gravitational wave signal.  The results also show that hierarchical black hole mergers \footnote{Multiple mergers of BHs that form more massive ones.}, could be observed in dense stellar systems, as shown in fig. \ref{fig:CollisionTree_BH-BH} and\ref{fig:ManyCollisionTree} (top panel), however as pointed out in recent studies \citep{Mapelli2020, Arca_Sedda2020} these events might be suppressed by the  gravitational wave kicks. The latter were not included in our simulations although we expect them to generate a large recoil velocity especially if the two colliding BHs have comparable masses  \citep{Campanelli2007, Baker2008, Kulier2015, Morawski2018, Zivancev2020}.

BH-IMBH coalescence, as well as the inspiral phase, will be detected by the next generation of gravitational wave detectors.  Events that involve an IMBH with mass $< 200$  \msun should be detected by LIGO  while signal generated by IMBHs with masses $< 2000$ \msun  should be observable with the Einstein Telescope \citep{Arca_Sedda2020}.

\section{Comparison with previous work}\label{section:comparison_with_other_work}



Various groups have predicted runaway merger scenario for the formation of VMSs and IMBHs in dense stellar environments.
For example, direct N-body simulations, carried out by \cite{Zwart2004}, indicate the formation of a few thousand solar masses stars produced by multiple stellar mergers. 
The simulated star clusters, containing $128k$ stars, were evolved for $12$ Myr using Starlab  \citep{Zwart2001} and  NBODY4 \citep{Aarseth1999}. The stellar evolution prescription adopted was based on \cite{Hurley2000}  for  stars with masses $\le 50$ \msun while more massive stars follow the evolution track given by  \cite{Stothers1997} and \citet{Ishii1999} .  With these models stars more massive than $260$ \msun collapse directly into IMBHs without losing mass in supernova explosions.

A similar mechanism is observed in the 30 simulations presented by \cite{Mapelli2016} where VMSs reach about $500$ \msun through runaway collisions. Each of these simulations is initialized with $10^5$ stars following a King density profile with \W$=9$ and \Rhb. The clusters were evolved for $17$ Myr using Starlab. Due to the different stellar evolution model adopted for massive stars, these simulations generate IMBHs of few hundred solar masses through direct collapse of VMSs (the VMSs at low metallicity lose a relatively small fraction of their masses).

The analysis of the $6000$ simulations of lower mass clusters presented in  \cite{DiCarlo2020} shows that BHs of about 300 \msun can form through dynamical interaction and collisions. The clusters, evolved using NBODY6++GPU, adopted the MOBSE stellar evolution \citep{Mapelli2017, Giacobbo2018}. This prescription is  based on \cite{Hurley2000, hurley2002} and it  includes new  prescriptions for massive stars reducing the mass loss in supernovae explosion and stellar winds.
The systems were initialized with an initial mass in the range between $10^3 $ and $3 \times 10^4$  \msun. The initial central densities and initial half-mass radii were computed as a function of the initial mass of the cluster.

Simulations evolved with Monte Carlo codes reveal similar outcomes. \cite{Freitag2006} computed over 100 models, varying the cluster size, particle number, and central concentration. They systematically changed the number of stars between $10^5$ and $10^8$ represented by a maximum of $9 \times 10^6$ particles. 
Their result show that 20 \% of the clusters with an initial central potential parameter \W  $\ge 8$ form a VMS with a mass $\ge 400$ \msun.
Other simulations carried out using Monte Carlo models show that multiple VMSs can form within the same cluster \citep{Gurkan2006}. 

The analysis of $2000$ of simulations \citep{Leigh2013, Giersz2015}, evolved using the MOCCA (MOnte Carlo Cluster simulAtor) code,  reveals the formation of IMBHs in dense stellar environment \citep{Giersz2015}. The outcome of these simulations indicates that about $20\%$ of the simulated clusters generate a BH with a mass larger than $100$ \msun. These BHs have formed through collisions between a VMS and a stellar BH. The formation path is very similar to the one indicated by our N-body simulations (see Figs. \ref{fig:ManyCollisionTree} and  \ref{fig:CollisionTree_IMBH}). However, in general, the IMBHs produced in MOCCA simulations tend to be systematically slightly more massive as the runaway main-sequence star collisions lead to more massive VMSs in MOCCA than in N-body.  In fact, in MOCCA simulations the stellar evolution time-step is performed at the end of the relaxation time-step (that is about 10 Myr). As a consequence of that, the masses of main-sequence stars, the mass segregation, and central density are larger in the MOCCA simulations than in N-body simulations leading to a larger interaction rate.

The analytical work carried on by \cite{Stone2017} shows how stellar mass BHs in nuclear star clusters can grow into IMBH through runaway tidal captures of low mass stars. As stated in their work, runaway tidal captures can be triggered only in massive and compact clusters with a velocity dispersions $\sigma > 40$ km/s. According to their criteria,
none of our models would have made IMBHs.  However, \cite{Stone2017} do not include massive stars and primordial binaries in their study, which are the key elements for the IMBH formation mechanism proposed in our work.

As we have shown in this section the debate whether and how IMBH form in dense star clusters is ongoing for at least two decades. Recent MOCCA Monte Carlo simulations have provided a wealth of data and answered the question positively. It is important to confirm MOCCA Monte Carlo results by direct N-body models, but the latter have suffered in the past from low statistical quality, if the particle number is small (say $10^4$ or less), and very demanding computing time requirements, if the particle number is large (e.g. $10^5$ or more). A strategy to balance low statistical quality of small N models has been to do larger samples of models and discuss their average \citep{GierszSp1994,DiCarlo2020}; but relevant astrophysical processes in star clusters (like two-body relaxation, close few body encounters, stellar evolution, tidal forces) do not scale with the same power of N. Therefore large sets of small N-body simulations can provide useful information to some degree, but can never fully substitute N-body simulations with more realistic larger particle numbers. \citet{baumgardt2017} present a very nice study using the method of small N samples and scaling to real star clusters; but still they are missing the effects of binaries and tidal fields, because they are difficult to scale. Our models exhibit IMBH formation in dense star clusters with an initially large particle number of more than 100k stars, $10\%$ of which are in binaries, and all relevant astrophysics; 80 such models were done using NBODY6++GPU for at least up to 300 Myr (8 models each for 10 different initial models). To our knowledge these are so far the largest direct N-body simulations of their kind, and in light of the discussion above they provide the so far strongest evidence for IMBH formation.

\section{Summary and Discussion}

We have provided evidence for the formation of intermediate mass black holes (IMBH) through collisions of massive stars, formation, and evolution of binaries including black holes. Debated for decades and recently underpinned by a large set of Monte Carlo (MOCCA) simulations our direct N-body models are the largest and longest simulations supporting this idea of IMBH formation in dense star clusters, made possible by the use of the massively parallel GPU accelerated code NBODY6++GPU and the use of suitable supercomputers in Germany and China.

We ran and analyzed $80$ N-body simulations of compact young massive star clusters with different central densities (central  potential parameters W$_0=6,7,8,9,10$) and sizes (half mass radii $R\mathrm{_{h}}=0.6, 1.0$ pc).
The simulated clusters were evolved for at least $300$ Myr \footnote{The simulations that formed an IMBH of about $350$ \msun were evolved for 500 Myr.}.
All our models lead to the collisional formation of at least one star above $100$ \msun (the upper initial mass function limit) and several simulations create stars with masses higher than $\sim 400$ \msun within the first $\sim 10$ Myr of cluster evolution. Most of the collisions were triggered by triple interactions between hard binaries and single objects. With the stellar evolution model assumed for this study, isolated massive stars cannot collapse directly into IMBHs (BHs with masses $>100$ \msun). Even stars with $\sim 500$ \msun lose most of their mass through stellar winds and collapse into a BH of about $30$ \msun.

However, a sizable fraction (about $20\%$) of our simulations result in the formation of IMBHs by means of direct collisions between stellar-mass BHs and massive stars as already observed in MOCCA simulations \citep{Giersz2015}. This process is more likely in compact clusters as they form more massive stars and it takes less time for the BHs to sink into the center. 
Nevertheless, if only a small fraction of the stellar mass is accreted in a collision with a BH (e.g. a collision fraction of \f = 0.1) the above process becomes unlikely for the formation of IMBHs in compact $\sim 7 \times 10^4 M_{\odot}$ clusters investigated in this study.

After its formation,  the IMBH  can still grow moderately colliding with other low mass BHs.
Here it is important to mention that kicks from gravitational radiation consequent to BH-BH mergers have not been implemented in our code. As a consequence of that, we might have overestimated the probability for the clusters to retain the IMBH because,  during a IMBH-BH collision,  the recoil velocity might exceed the escape velocity \citep{Campanelli2007, 
Baker2008, Kulier2015, Morawski2018, Zivancev2020}.  

During the growth process of all simulations with IMBHs,  with one exception shown in Fig. \ref{fig:CollisionTree_BH-BH}, there are no BH - BH mergers\footnote{In general, BH-BH collisions events are rare. This type of binaries must undergo several strong close interactions to enter the post-Newtonian regime where gravitational radiations can lead to a rapid coalescence. Because of these interactions, the binary is often ejected from the cluster before the merger occurs.} before the formation of the IMBH in a VMS - stellar BH collision. Therefore the inclusion of gravitational kicks will not change this result. After the IMBH has formed it occasionally collides with a stellar mass BH in an intermediate mass-ratio inspiral (IMRIs) event, which has the potential to kick the IMBH out of the cluster. If common, such a process might explain the missing observational evidence for IMBHs in present day globular clusters. IMBHs might have formed in many GCs early on and, once lost, float around in the galaxies.

Adding gravitational wave kicks and spins will be possible in the future using the approximate models of \citet{Baker2008} for the kick velocity (magnitude and direction) and a new model of how initial BH spins depend on mass and metallicity by \citet{Belczynski2017}. \citet{Morawski2018} have analyzed large samples of BH mergers from MOCCA simulations, and show that the BH retention fraction in the cluster varies between 20\% and 100\% depending on evolutionary time and parameters of the cluster. \citet{Brem2013} have included full Post-Newtonian dynamics in their N-body simulation and reproduced results of \citet{Rezzolla2008}, who fitted fully relativistic models. This could also be used to derive recoil velocities, in the way done by \citet{Gerosa2018}, the latter again using fully relativistic modeling.

Our results show that the models R06W6 and R06W9, despite the difference in the initial central density,  have a comparable probability to form an IMBH.
The different evolution of the inner part of the clusters in these two models seems to mitigate the impact of the initial central density on the probability to form an IMBH: in very concentrated systems, the high central density forces the clusters to expand because of early energy generation by primordial binaries; on the other hand, less dense clusters undergo core collapse. Therefore, already at the very beginning of the simulation, the initial difference in central concentration between the models is reduced.

Our results indicate that compact star clusters can rapidly generate an IMBH of few hundred solar masses in about $5-15$ Myr.  
Assuming the scenario that nuclear star clusters are generated by globular clusters that spiral toward the nucleus \citep{Tremaine1975},  the IMBHs, if present in the clusters,  can further grow in mass colliding with each other and swallowing smaller objects in the center of the nuclear cluster  \citep{Arca-Sedda2018, Arca-Sedda2019, Askar2020} leading to the formation of a SMBH.
This scenario is investigated by  \cite{Stone2017} adopting an analytical approach. They show how low mass BHs located in dense nuclear star cluster could rapidly grow in mass via runaway tidal captures,  transforming the cluster into a SMBH.

With order $10^5$ particles our models are currently the best available (in the sense of modeling all processes directly for the simulated particle number, without any scaling or averaging). However, young massive clusters in our galaxy and massive extragalactic clusters (also: nuclear star clusters) can be much more massive with particle numbers of up to $10^8$ or more. NBODY6++GPU has been used for the million-body DRAGON simulations \citep{Wang2015,Wang2016} and for million body simulation of a nuclear star cluster \citep{panamarev2019}. But in the first case, the central density was much lower than in this paper, so the DRAGON simulations are not prone to IMBH formation, and in the second case still, some scaling had to be used, because $10^6$ particles are not enough to model nuclear star clusters. In the next ongoing studies, we are running dense star cluster models with a million bodies, and more in the future. It is feasible because only a shorter simulation time is needed (a few hundred Myr versus 12 Gyr for DRAGON). This will help us to get a better understanding of the statistics of the presence of IMBHs, not only in our Galaxy but also out to distant regions relevant for LIGO/Virgo and space-based gravitational wave detections.

In future models we plan to improve the modeling of the external tidal forces on the cluster, reflecting its true orbit around the host galaxy; also stellar evolution has been significantly updated recently\footnote{The most updated version of NBODY6++GPU (partly inspired by LIGO data) contains some recent stellar evolution updates  \cite[see][]{Belczynski2008, Banerjee2020} These updates were made after the completion of the simulations presented in this work.} and is used for our ongoing and future simulations, see for information e.g. \citet{banerjee2019}.



\section*{Acknowledgements}

 We acknowledge Sambaran Banerjee for the updated BSE code release before publication. FR thanks Dali University, Yunnan, China and Zhongmu Li for kind hospitality and support during a workshop in Dali in 2019 as well as National Astronomical Observatory, Chinese Academy of Sciences, Silk Road Project for hospitality during a research visits in Beijing. This work has been partly supported by Sino-German cooperation (DFG, NSFC) under project number GZ 1284, and by Chinese National Science Foundation under grant No. 11673032 (RS). TN acknowledges support from the Deutsche Forschungsgemeinschaft (DFG, German Research Foundation) under Germany's Excellence Strategy - EXC-2094 - 390783311 from the DFG Cluster of Excellence "ORIGINS".

The authors gratefully acknowledge the Gauss Centre for Supercomputing (GSC) e.V. (www.gauss-centre.eu) for funding this project by providing computing
time through the John von Neumann Institute for Computing (NIC) on the GCS Supercomputers JURECA and JUWELS at Jülich Supercomputing Centre (JSC).

FR and RS thank Hyung Mok Lee and the Korea Astronomy and Space Science Institute (KASI) in Daejoen, Korea (Rep.), for financial support during the KCK11 meeting in Dec. 2019.

FR acknowledge with pleasure the  helpful discussion with Francesco Flammini, Manuel Arca Sedda and Agostino Leveque.

MG was partially supported by the Polish National Science Center (NCN) through the grant UMO-2016/23/B/ST9/02732.

PB acknowledges support by the Chinese Academy of Sciences through the Silk Road Project at NAOC, the President’s International Fellowship (PIFI) for Visiting Scientists program of CAS, 
the National Science Foundation of China under grant No. 11673032.

NCS received financial support from NASA, through the NASA Astrophysics Theory Research Program (Grant NNX17AK43G; PI B. Metzger). He also received support from the Israel Science Foundation (Individual Research Grant 2565/19).

This work was supported by the Deutsche Forschungsgemeinschaft (DFG, German Research Foundation) – Project-ID 138713538 – SFB 881 (‘The Milky Way System’), by the Volkswagen Foundation under the Trilateral Partnerships grants No. 90411 and 97778.

The work of PB was also partially supported under the special program of the NAS of Ukraine ‘Support for the development of priority fields of scientific research’ (CPCEL 6541230).

\addcontentsline{toc}{section}{Acknowledgements}




\bibliographystyle{mnras}
\bibliography{references} 



\end{document}